\documentclass[onecolumn]{article}
\usepackage[a4paper, total={6in, 8in}]{geometry}
\usepackage{natbib}
\setcitestyle{numbers} 
 \usepackage{authblk} 
\usepackage{amssymb}
\usepackage{multirow}
\usepackage{lscape}
\usepackage{color}
\usepackage{tabularx}
\usepackage{url}
\usepackage{dirtytalk}
\usepackage{xr}
\usepackage{fancyvrb}
\usepackage{fvextra}
\usepackage{csquotes}
\usepackage{longtable}
\usepackage{amsmath}
\usepackage{hyperref}
\usepackage{booktabs}
\usepackage{listings}
\usepackage{environ}
\usepackage[edges]{forest}
\usepackage{fontawesome5}
\usepackage{adjustbox}
\usepackage{tikz}
\usetikzlibrary{arrows,
	chains,
	positioning,
	shapes.geometric
}
\usepackage[skins]{tcolorbox}

\usepackage[textsize=small]{todonotes}

\usepackage{colortbl}
\usepackage{enumitem}

	\newcolumntype{L}[1]{>{\raggedright\arraybackslash}p{#1}}

\definecolor{gr}{gray}{0.9}

\newcommand{\smellinlinefield}[2]{\par\vspace{0.2em}{\small\scshape\color{black!80} #1}\hspace{0.8em}#2\par
}

\newcommand{\smellblockfield}[1]{\par\vspace{0.2em}{\small\scshape\color{black!80} #1}\par\nopagebreak\vspace{-0.3em}
}

\newcommand{\smelltitle}[2]{\par\vspace{1.2em}\noindent\textbf{#1: #2}\par\nopagebreak
	\vspace{-0.2em}}

\newcommand{\scope}[1]{\smellinlinefield{Scope of Impact}{#1}}

\newcommand{\descri}{\smellblockfield{Description}}

\newcommand{\implications}{\smellblockfield{Implications}}

\newcommand{\causes}{\smellblockfield{Causes}}

\newcommand{\facets}{\smellblockfield{Facets and Implications}}

\newcommand{\references}[1]{\smellinlinefield{References}{#1}}

\newcommand{\example}{\smellblockfield{Example}}

\newenvironment{smellfacets}{\par\vspace{0.4em}\noindent
	\tabularx{\linewidth}{@{} l @{\hspace{0.7em}} X @{}}
}{\endtabularx\par\vspace{0.4em}}

\newcommand{\facet}[2]{#1 & #2 \tabularnewline[0.0em] }

\newcommand{\impact}[1]{& \noindent\makebox[1.2em][l]{\textcolor{gray!70}{$\rightarrow$}}\parbox[t]{\dimexpr\linewidth-1.2em\relax}{#1}\tabularnewline[0.8em] }

\usepackage{xcolor}

\definecolor{diffred}{RGB}{255, 220, 220}
\definecolor{diffgreen}{RGB}{220, 255, 220}

\lstdefinelanguage{diff}{
	basicstyle=\ttfamily\small,
	columns=fullflexible,
	keepspaces=true,
	breaklines=true,
	morecomment=[f][\color{red!70!black}]{-},
	morecomment=[f][\color{green!50!black}]{+},
}

\title{Towards a Taxonomy of Software Log Smells}

\author[1]{Nyyti Saarimäki}
\author[1]{Zahra G. Yazdi}
\author[2]{Donghwan Shin}
\author[1]{Domenico Bianculli}

\affil[1]{University of Luxembourg, Luxembourg, \texttt{firstname.lastname@uni.lu}}
\affil[2]{University of Sheffield, United Kingdom, \texttt{d.shin@sheffield.ac.uk}}
\date{}

\begin{document}
	
\maketitle

\begin{abstract}

	\textbf{Context:} Logging is an important part of modern software projects; logs are used in several tasks such as debugging and testing. Due to the complex nature of logging, it remains a difficult task with several pitfalls that could have serious consequences. Several other domains of software engineering have mitigated such threats by identifying the early signs of more serious issues, i.e., ``smells''. However, in the context of software logging, the concept of smells has only been investigated in a limited and narrow scope.
	
	\noindent\textbf{Objective:}  The goal of this study is to systematically organize logging issues reported in the literature into a taxonomy of log smells that can help  practitioners to better understand logging quality issues. To further help the practitioners and to identify issues that need more attention from the research community, we also map the identified smells to existing tools repairing them.
	
	\noindent\textbf{Methods:} We conducted a survey of the scientific literature to identify logging issues and related tools. After extracting relevant data from 61 articles, we used open coding to define logging issues
	and applied card sorting  to derive log smells from these issues. Finally, we classified the tools based on their reported output.
	
	\noindent\textbf{Results:} We present a taxonomy of nine log
	smells, describe several facets for each of them, and report the results of a preliminary validation of the taxonomy using a dataset of logging issues extracted from open-source projects.
        We also review existing tools repairing/removing some of these facets, highlighting the lack
	of tools addressing some log smells and identifying future
	research opportunities to close this gap.
	
	\noindent\textbf{Conclusions:} Our findings indicate that software logging is susceptible to a wide variety of log smells affecting both the logging code and the resulting log files. Understanding these smells, along with their causes and consequences, can help reduce their occurrence and improve logging quality.

\end{abstract}

\section{Introduction}
\label{sec:intro}

Logging is a critical part of modern software development, and it is used in tasks such as anomaly detection and quality assurance~\citep{candido_log-based_2021}. However, producing good logs is non-trivial; it requires making technical and design decisions as
well as coordinating the collaboration among developers. Good quality logging is important as poor logging can cause serious issues and delays in a software project~\citep{he_empirical_2022}.

Due to the importance of logging, a wide variety of issues and practices related to software logging have been studied in the literature. Prior work has investigated aspects related to, among others, system performance, logging levels, security, maintainability, and the placement of logging statements in source code. To organize this broad body of knowledge, researchers have proposed categorizations from a variety of perspectives, including log instrumentation~\cite{chen_survey_2022}, log analysis and parsing~\cite{he2021survey}, and the decisions associated with what, where, when, and how to log~\cite{gu_logging_2023}.

While these categorizations are useful for understanding logging practices and which aspect of logging is involved, they are not designed to organize logging issues themselves. The \textit{what, where, when, and how to log} framework is a common way to classify logging concerns in the literature. However, the categories may group together issues with very different characteristics, causes, and implications. For example,~\citet{gu_logging_2023} classify ``lacking crucial messages'', ``redundant or useless messages'', ``incorrect or ambiguous messages'', and ``leakage of sensitive data'' under the ``what to log'' category.  Although these issues all relate to the information being logged, each represents a distinct quality concern that can manifest itself in multiple ways. Therefore, there is a need for a complementary perspective that focuses on logging quality problems rather than on the decisions involved in producing logs.

In other areas of software engineering (e.g., requirements engineering, coding, and testing), indicators of deeper design problems, recurring problems, or issues impacting quality are commonly referred to as \emph{smells}~\citep{sharma_survey_2018}. The concept has also been adopted in the context of logging. However, to the best of our knowledge, only two studies have explicitly defined log smells. \citet{chen_characterizing_2017} define logging code smells as ``poor design and implementation choices when developing the logging code'', while \citet{li_dlfinder_2019} describe duplicate logging code smells as ``surface indication that usually corresponds to a deeper problem in the system''.

However, these works present only a limited subset of log
smells. \citet{chen_characterizing_2017} discuss logging code smells
only briefly through two duplication-related anti-patterns, while
\citet{li_dlfinder_2019} focus exclusively on duplicate logging code. Consequently, existing work establishes the notion of log smells but does not provide a broader understanding of the different types of log smells discussed in the literature. Therefore, there is a need to build on prior work and investigate log smells more generally.

The goal of this study is to build on existing definitions of log smells and systematically organize reported logging issues into a taxonomy of log smells, and to understand whether the research community has already proposed solutions to repair them. The taxonomy will help
practitioners become more aware of logging issues and their
consequences. Identifying potential logging smells at an early stage
helps avoid more serious issues. Furthermore, mapping the available tools to the log smells identified in 
the taxonomy will help practitioners select and apply appropriate tools to avoid such smells in their code or logs. 
Such a mapping will also pinpoint smells that are not (fully) addressed by current mitigation
tools, providing directions for future research work.

We conducted the study as a literature survey from Google Scholar, IEEE Xplore, Scopus and ACM Digital Library. We extracted problems related to logging highlighted in the literature
and grouped similar problems together using open coding to form
logging issues. The issues were categorized into groups by conducting
several rounds of card sorts and discussions among the authors. We
categorized the tools according to their output.
We derived a taxonomy comprising nine log
smells, which can be further divided into several facets.
To assess the applicability of the taxonomy beyond the surveyed literature, we additionally conducted a preliminary validation using a dataset of logging issues extracted from open-source projects.
The process of defining the log smells in the catalog also revealed five direct
causes and four consequences associated with these smells. We identified 16 tools that
address (and repair) facets from eight log smells.

In summary, the main contributions of this article are as follows:
\begin{itemize}
	\item Building on existing definitions of log smells, we derive a taxonomy containing nine log smells and their facets.
	\item Mapping log smells and their facets to tools that repair them.
\end{itemize}

The remainder of the paper is structured as follows.
Section~\ref{sec:preliminaries} provides background information and definitions on
logging and smells. Section~\ref{sec:method} illustrates the study design. Section~\ref{sec:results} presents the results, while Section~\ref{sec:discussion} discusses them. Section~\ref{sec:threats_to_validity} discusses the threats to
validity. Section~\ref{sec:related_work} provides an overview of the
related work. Finally, Section~\ref{sec:conclusion} concludes the paper.

 \section{Backgroung \& Definitions}
\label{sec:preliminaries}

\subsection{Logs}

A \textit{log} is a sequence of log entries that are usually saved in a
text file called a \textit{log file}. A \textit{log entry} is either a
single line or multiple contiguous lines of text consisting of a
(single) header and a message. The \textit{header} provides the practitioner
meta-knowledge about the entry, such as the time the logging code was
executed (timestamp), the identifier for the process or component that
generated the log entry (process id), and the severity of the logged
event (logging level). The header format is typically defined by the
underlying logging framework. The \textit{log message} is part of
a log entry that presents information about the state of the software
system at runtime. It consists of a static part (often called a \emph{log (message) template}) corresponding to the predefined text written by
the developer as part of a logging statement and a dynamic part,
which represents the values of the variables captured in the log
message at a certain moment of the software execution. An example of a log entry that highlights its different elements is shown below:

{
	\footnotesize
	\DefineShortVerb{\|}
	\[
	\begin{split}
		&\underbrace{\verb|2024-08-05 11:54:12|}_{\text{Timestamp}} \underbrace{\verb|ERROR|}_{\text{Logging level}} \underbrace{\verb|[database]|}_{\text{Process id}} \  \underbrace{\verb|Failed to connect to database:|}_{\text{Static part}} \\ 
		&\hookrightarrow
		\underbrace{\verb|Connection timeout|}_{\text{Dynamic part
				(variable)}}
		\underbrace{\verb|. Abort proc.|}_{\text{Static part}}
	\end{split}
	\]
}  

Log entries (and the log file) are produced by \emph{logging
	code}. An example of logging code generating the example log entry
shown above is given in Listing~\ref{lst:logging_code_example}. Often, the code is tangled with the feature code and, hence, scattered throughout the project~\citep{Kiczales1997}. The logs and logging code have a tight relationship as the logging code produces the logs. Therefore, some characteristics between the two are shared while others are individual, and the logging and feature code are susceptible to similar issues. For example, if the logging code has an incorrect log message, this is also reflected in the log entries produced by the corresponding statement. However, if the logging code statement is unreasonably long or difficult to understand, that is a problem only in the code.

\begin{lstlisting}[language=Python, caption=An example of logging code., label=lst:logging_code_example, basicstyle=\footnotesize\sf, showstringspaces=false]
try:
     logger.info(f'Connecting to database')
     conn = sqlite3.connect(db_file)
except sqlite3.Error as e:
    logger.error(f'Failed to connect to database: {e}. Abort proc.')
\end{lstlisting}

Logs can be inspected differently depending on whether the source code that produces the logs can be accessed. A \textit{black-box setting}
occurs when only the log files are accessible. Consequently, it is impossible to exactly know how and where the log entries were produced and, in such a case, having ``good quality'' log files is crucial. However, in a \textit{white-box setting}, the source code and log files are available, and it is possible to understand which part of the source code produced which log entry.

\subsection{Software Smells}
\label{sec:bg-software-smells}

The term ``smell'' in software engineering was
popularized by Beck and Fowler through their characterization of code
smells~\citep{fowler1999}; since then,  it has been widely adopted across the field. In their survey, \citet{sharma_survey_2018} identified 14 groups of smells, including achitecture~\citep{brown_antipatterns_1998} and test smells~\citep{van2001refactoring}.
More recently, additional categories have emerged, such as bug tracking smells \cite{QAMAR2022106972} and code review smells \cite{dogan_towards_2022}.

\citet{sharma_survey_2018} observed that the term ``smell'' is typically used to denote indicators of deeper design issues, poor solutions, recurring problems, quality-impacting problems, or violations of best practices. Hence, smells do not prevent software from functioning correctly, but they degrade quality and may lead to more serious issues over time.
Although smell and antipattern are sometimes used interchangeably, they differ conceptually. Antipatterns are intentional but misguided solutions that create negative consequences~\citep{koenig1998patterns}, whereas smells signal potential problems and are not necessarily introduced deliberately~\citep{sharma_survey_2018}.

\subsection{Log Smells}
\label{sec:bg-log-smells}

In this paper, we extend the previously presented definitions~\cite{chen_characterizing_2017, li_dlfinder_2019}  of a log smell to have the following definition:
\begin{tcolorbox}
	\begin{displayquote}
		Log smell is a poor design choice or an issue impacting quality that could result in a more serious problem, affecting the logging code, the log file, or both.
	\end{displayquote}    
\end{tcolorbox}

The definition has two main aspects. 
First, a smell is an indication of a deeper problem, but is not a serious problem itself. 
For example, a missing logging statement in a catch-block does not prevent the system from running; 
however, when looking at the code, it is usually clear that a logging statement is missing. 
Second, log smells affect not only the logging code, but also the
log files. This is a natural consequence of the fact that the logs are
generated by the logging code. However, the two are not the same, and
not all issues affecting the code affect the log files and vice
versa. The distinction between the two is especially important to
deal with cases in which the engineer reading the logs does not have access to the code that created the log or documentation of the used logging practices. 

Each smell can manifest itself in several different ways; in this paper, we refer to them using the term \textit{facet}. For example, having wrong static messages and messages in which the grammar is incorrect affect both the quality of static messages, but they create an issue in different ways. Therefore, they would be considered as facets, while in general, having issues with static messages would be a log smell. 

The two prior works that define log smells are discussed in detail in Section~\ref{sec:rw_log_smells}.

 \section{Study Design}
\label{sec:method}

The goal of this study is to identify and categorize software log smells, as well as to identify tools to remove or repair them.
Based on this goal, we define the following research questions:

\begin{description}
	\item[RQ1.] \emph{What logging issues are log smells or directly related to them?} 
\item[RQ2.] \emph{Which log smells have tools or other automated techniques able to remove or repair them?}
\end{description}

Although the literature has investigated a wide variety of issues related to logging, only two works have explicitly defined the concept of log smells~\cite{chen_characterizing_2017, li_dlfinder_2019}. However, to the best of our knowledge, no previous work has systematically synthesized the reported logging issues into a taxonomy that classifies log smells as well as their causes and consequences. Building on the existing definitions of log smells, we address RQ1 by analyzing logging issues reported in the literature and investigating their relationship to log smells.

The answer to RQ2 aims to understand which log smells have a
tool or other automated solution able to remove/repair a certain smell.
By mapping the tools identified in the literature survey to the
different aspects of logging issues, we can, on the one hand, provide
practitioners with a catalog of tools for removing/repairing log
smells; on the other hand, we can highlight areas in which limited tool support was identified, and that may deserve additional investigation.
We note that RQ2 focuses on the \emph{removal} or \emph{repair} of log smells rather than their \emph{detection}, which has already been investigated by \citet{MADI2026107961} (see Section~\ref{sec:RW_SLRs}).

\subsection{Taxonomy Creation}
\label{sec:taxonomy_creation}

We created the taxonomy following the method presented
by~\citet{usman_taxonomies_2017}. The method has four phases
(planning, identification and extraction, design and construction, validation)
consisting of a total of 13 steps. In this section, we describe our decisions and actions for each step. 

\vspace{\baselineskip}
\noindent\textbf{Planning.} Establish the foundation for designing the taxonomy.

\begin{enumerate}
	\setlength\itemsep{0em}

	\item \textit{Knowledge area:} Logging is widely used across software engineering, so we designed the taxonomy to span the entire field rather than focus on a single domain. Therefore, we do not present issues that are specific to a single downstream task.
	\item \textit{Goal:} The taxonomy aims to classify log smells based on the logging aspects they impact, using aspects general enough to align with the knowledge area.
	\item \textit{Subject matter:} This defines the unit of classification, which in this taxonomy is the log smell.
	\item \textit{Classification structure type:} The taxonomy’s structure follows from the characteristics of the subject matter. Since each log smell can be uniquely grouped according to the logging aspect it influences, a hierarchical structure with a single root class was selected.
	\item \textit{Classification procedure type:} Since the taxonomy aims to provide a qualitative characterization of log smells, its classification procedure type is also qualitative.
	\item \textit{Identify information sources:}  The data collection method for this work was a literature survey, which is described in detail in Section~\ref{sec:method_survey}.
\end{enumerate}

\noindent\textbf{Identification and extraction.} Extract terms and unify the terminology.

\begin{enumerate}[resume]
	\setlength\itemsep{0em}
	\item \textit{Extract terms:} We collected relevant terms and issue descriptions from the scientific literature as part of the literature survey.
	\item \textit{Terminology control:} The terms were categorized into groups, each describing an issue in logging. Once the final log smells were formed, we created names for them; this process is described in Section~\ref{sec:data_analysis}.
\end{enumerate}

\noindent\textbf{Design and construction.} Identify and describe the taxonomy’s dimensions, categories, and their relationships.

\begin{enumerate}[resume]
	\setlength\itemsep{0em}
	\item \textit{Identify dimensions:} Hierarchical taxonomies have only one dimension. In our case, the top-level dimension is log smell.
	\item \textit{Identify categories:} The categories were
	defined bottom-up, i.e., we started creating them from the log smells themselves.  We defined two second-level dimensions (log artifact, logging code), four third-level dimensions (log file, log entry content, transparent to the logs, and invisible to the logs), and nine fourth-level dimensions. The last level is the actual log smells; these are described in Section~\ref{sec:log_smells}.
	\item \textit{Identify relationships:} The taxonomy is organized as a hierarchy, with its structure outlining the relationships.
	\item \textit{Guidelines for usage and updating:} The proposed taxonomy can be further developed by extending the literature review or conducting a survey with developers. The taxonomy is applied by determining which aspect of logging is primarily affected by the log smell. If an issue that qualifies as a smell does not fit any existing category in the taxonomy, the taxonomy should be expanded. It may also be the case that we have not labeled the issue as a smell, but instead as a cause, a consequence, or another type of logging-related issue.
\end{enumerate}

\noindent\textbf{Validation.} Define any validation conducted on the taxonomy.
\begin{enumerate}[resume]
\item To assess whether the proposed taxonomy captures logging issues occurring in practice, we conducted a preliminary validation using the dataset provided by~\citet{Wang2025}. We selected this dataset because it contains manually validated real-world logging issues extracted from mature open-source systems. The dataset contains 164 manually validated logging code issues extracted from Hadoop, HBase, Hive, YARN, and ActiveMQ using their issue-tracking systems and commit histories.

Authors~1 and~2 independently classified each issue according to the proposed log smell taxonomy. Specifically, the authors assigned each issue to one or more of the identified log smells and, where applicable, to a corresponding facet. Issues that could not be represented by any of the identified smells were labeled as not covered by the taxonomy. The classification was performed based on the commit message and the associated code change, and importantly, we did not use the defect categories provided in the dataset. The authors subsequently reviewed the classifications together, corrected labeling mistakes, and discussed disagreements to reach a consensus classification. When consensus could not be reached, Authors~3 and~4 were consulted. We report inter-rater agreement using Cohen's Kappa ($\kappa$).
	
\end{enumerate}

\subsection{Literature Survey}
\label{sec:method_survey}

We answered our RQs by surveying existing papers on software logging. An overview of the study method is presented in Figure~\ref{fig:survey_overview}. We conducted a literature survey rather than a full systematic literature review (SLR). While we did not perform all activities typically associated with an SLR, such as snowballing, we aimed to follow SLR best practices and guidelines wherever possible~\cite{kitchenham_segress_2023}. The survey was conducted by Authors 1 and 2, while all authors participated in discussions throughout the process to reduce potential bias. We provide further details about the process in the replication package.

\begin{figure}
	\centering
	\begin{adjustbox}{width=1.15\textwidth}
	\begin{tikzpicture}[auto,
	block_center/.style ={rectangle, draw=black, thick, fill=white,
		text width=8em, text centered,
		minimum height=4em},
	block_left/.style ={rectangle, draw=black, thick, fill=white,
		text width=16em, text ragged, minimum height=4em, inner sep=6pt},
	block_noborder/.style ={rectangle, draw=none, thick, fill=none,
		text width=18em, minimum height=1em, text depth = 0.6 cm},
	line/.style ={draw, thick, -latex', shorten >=0pt},
	disc/.style = {shape=cylinder, draw, shape aspect=0.2, shape border rotate=90, align=center, text width=8em, minimum height=3em},
	doc/.style={document, draw, minimum width=21mm, minimum height=16mm, text width=8em, text centered}]
	\footnotesize
	
	\matrix [column sep=-18mm, row sep=4mm] {
		\node [left, disc] (referred) {4 \\ Databases }; & 
		\node[left, block_center] (criteria) {Inclusion and exclusion criteria}; &
		\node[left, block_center] (extraction) {Data extraction }; &
		\node[left, block_center] (analysis) {Analysis}; &
		\node[left, block_center] (results) {Results}; \\
		\node [block_noborder] (excluded1) {Search: \\ - 14 search queries\\ - 475 search results \\
		- 377 unique results}; & 
		\node [block_noborder] (excluded2) {Excluded (316 papers): \\
			- Inclusion criteria not \\
			  \hspace{0.3em} met (300 papers) \\
			- Exclusion criteria met \\
			\hspace{0.3em} (16 papers)}; 
		& \node[block_noborder] (included_papers) {From 61 included papers:\\ - Reported logging issues\\ - Logging related tools};
		& \node[block_noborder] (analysis_techniques){Techniques:\\ - Open coding \\ - Open and closed card \\ \hspace{0.3em} sort}; 
		& \node[block_noborder] (results_rq){Main results: \\ - Log smell taxonomy \\ - Mapping of existing \\ \hspace{0.3em}  tools with log smells}; \\
	};
	\begin{scope}[every path/.style=line]
\path (referred) -- (criteria);
		\path (criteria) -- (extraction);
\path (extraction) -- (analysis);
\path (analysis) -- (results);
\end{scope}
\end{tikzpicture}
\end{adjustbox}
 	\caption{Overview of the study method.}
	\label{fig:survey_overview}
\end{figure}

\noindent\textbf{Information sources.} 
We searched Google Scholar (GS), Scopus, the ACM Digital Library, and IEEE Xplore.

\noindent\textbf{Search strategy.} 
We used  the 14 search queries presented in Table~\ref{tab:search_strings} and applied them to all databases. The search queries were designed to capture the terminology commonly used when discussing logging issues and logging quality. Extremely general queries were avoided because they produced large numbers of irrelevant non‑software results. We did not apply snowballing, as the goal was to investigate log smells rather than conduct a full SLR.

Because Google Scholar rankings change over time, we inspected the top
results and report them in two separate columns in
Table~\ref{tab:search_strings}. The first corresponds to the initial
survey conducted in early 2024; the second reflects an extension conducted in fall 2025, which included only papers published after the initial search.

\begin{table*}
	\centering
	\small 
	\caption{Search strings and number of results for the different databases. Note that some papers were included in the results of several queries.}
	\label{tab:search_strings}
       	\begin{tabular}{p{7.1cm}ccccc}
		\toprule
		\textbf{Query}	&	\textbf{ACM}	&	\textbf{Scopus}	&	\textbf{IEEE}	&	\textbf{Scholar}$^*$	&	\textbf{Scholar}$^*$ \\
					&		&		&		&		&	\textbf{24-25}$^*$ \\
		\midrule

``log quality'' AND software	&	27	&	29	&	5	&	2840	&	7	\\
``logging quality'' AND software 	&	17	&	5	&	0	&	249	&	12	\\
software AND (``log issues'' OR ``log issue'')	&	24	&	7	&	1	&	684	&	6	\\
``logging code smell''	&	2	&	3	&	1	&	5	&	0	\\
``bad logging practices'' AND software	&	1	&	1	&	1	&	16	&	0	\\
``bad logging practice'' AND software	&	1	&	1	&	0	&	10	&	0	\\
``software log quality''	&	0	&	0	&	0	&	0	&	1	\\
``software engineering'' AND ( ``logging issues'' OR ``logging issue'')	&	16	&	3	&	2	&	132	&	11	\\
``log smell''	&	0	&	0	&	0	&	9	&	0	\\
``log smell'' AND software	&	0	&	0	&	0	&	1	&	0	\\
``logging smell'' OR ``logging smells''	&	1	&	0	&	0	&	3	&	1	\\
``log anti-pattern'' OR ``log anti-patterns''	&	1	&	0	&	0	&	1	&	1	\\
``logging anti-pattern'' OR ``logging anti-patterns''	&	6	&	0	&	0	&	26	&	6	\\
``logging bad practices'' AND software	&	0	&	0	&	0	&	0	&	1	\\
\midrule
\textbf{Total}	&	96	&	49	&	10	&	3976	&	46	\\
		\bottomrule
		\multicolumn{6}{l}{* Results after the first five pages ($\approx$50 results) were only browsed until page 20 ($\approx$200 results).} \\
	\end{tabular}
\end{table*} 
\noindent\textbf{Eligibility criteria}. Inclusion criteria are denoted as (I) and exclusion criteria as (E).
We included peer‑reviewed papers written in English (I1), published in 2000 or later (I2), and related to software logging (I3). Because our study is qualitative and aims to capture as many logging issues as possible, we did not impose additional quality criteria such as publication venue or research methodology.
Secondary studies were included when they synthesized logging issues or listed relevant tools.

We excluded papers that had an extension paper among the other included papers and kept only the most recent version of a paper (E1). 
Finally, papers that did not describe issues related to software logging were excluded (E2).

\noindent\textbf{Selection and data collection process.} 
From the 377 unique papers identified through the search process, 61 were included in the survey. To assess the reliability of the study selection process, Authors 1 and 2 independently screened all candidate papers and applied the inclusion and exclusion criteria. The authors disagreed on 21 papers; 16 disagreements were resolved through discussion, while the remaining five were resolved with the involvement of Authors 3 and 4. The resulting Cohen's Kappa value was 0.807, indicating substantial to almost perfect agreement~\cite{Landis1977}.

From each of the 61 included papers, we collected metadata (authors, title, year, venue) and extracted (i) reported logging issues for RQ1 and (ii) logging-related tools for RQ2. Issues were recorded primarily as direct quotes. Tools were recorded under the authors’ given names; when unnamed, we assigned an identifier based on the first author’s surname and publication year.

\subsection{Data Analysis}
\label{sec:data_analysis}

\noindent\textbf{RQ1. Identifying and categorizing log smells.}
We analyzed the extracted data in two steps using qualitative methods. The two-step approach allowed us to explore different ways to categorize all potential logging issues initially identified in the data. 

\medskip

\noindent\textit{Step 1: Identifying issue categories.}
Author 1 applied open coding~\cite{stol_grounded_2015} to consolidate the extracted logging issues, after which all authors participated in an open card sort~\cite{spencer_card_2009} to explore alternative groupings. Before performing the sort, the authors agreed that the categorization should reflect logging issues rather than logging processes; for example, existing frameworks such as ``why, where, what, and how to log'' were not appropriate for this purpose~\cite{gu_logging_2023}.

Authors reviewed and discussed the resulting groupings and refined issue descriptions accordingly. This process led to an initial set of categories relevant to RQ1: causes of log smells, log smells, and consequences of log smells. Specifically, we defined the categories as follows:
\begin{itemize}
	\item \textit{Causes of log smells}: Issues that can lead to log smells. The issues in the group are not necessarily directly visible in the logging code or log files, but they are aspects of the project that, if not managed properly, will lead to log smells.
	\item \textit{Log smells:} Issues in the logging code or the actual logs that might cause problems but do not affect the functionality of the project. Smells are generally considered  indicators of underlying technical debt. A more detailed description of log smells is given in Section~\ref{sec:bg-log-smells}. 
	\item \textit{Consequences of log smells:} Issues caused by log smells that manifest in the logging code, log files, or logging system. We consider only direct issues
	caused by the log smells and exclude issues that are affected but not directly caused by them, such as logs that are useless for failure analysis.
\end{itemize}

Additionally, we used a fourth category called ``Something else'' that caught all issues not directly related to log smells. 

\medskip
\noindent\textit{Step 2: Classifying logging issues.}
We then performed a closed card sort  in which Authors 1, 3, and 4 classified the 51 identified logging issues into the four predefined categories. The authors were unanimous on 12 issues (23\%), two authors agreed on 34 issues (67\%), and five issues received three different classifications (10\%). Consequently, 46 of the 51 issues (90\%) received a majority classification, meaning at least two authors assigned it the same category, before discussion. We additionally calculated Fleiss' Kappa and obtained a value of 0.244.

The apparent discrepancy between the high proportion of majority classifications and the relatively low Kappa value results from the different aspects of agreement captured by these measures. The majority classification metric considers an issue successfully classified whenever at least two of the three authors selected the same category. In contrast, Fleiss' Kappa gives greater weight to unanimous agreement among all raters and adjusts observed agreement for agreement expected by chance. Consequently, issues with majority agreement but not unanimous agreement contribute less to the Kappa statistic than to the majority-classification measure. Nevertheless, the obtained value indicates only fair agreement~\cite{Landis1977}, showing that the independent classifications often differed before discussion. Therefore, the card sort results were not used as the final classification, as explained below.

Although Fleiss' Kappa provides a useful indication of the
  level of independent agreement, we emphasize that the objective of
  the card sorting was neither to produce the final classification nor
  to maximize inter-rater agreement. Rather, the exercise served as a structured starting point for the taxonomy development by identifying both areas of agreement and cases requiring further discussion. The fact that 90\% of the issues received a majority classification suggests that the authors shared a broadly similar understanding of the four categories prior to discussion and that the card sort successfully established a common basis for the subsequent taxonomy development discussions. 
	
After the closed card sort, all issues, including those with majority agreement, 
were reviewed and discussed until consensus was reached. During these discussions, 
we merged overlapping issues, clarified definitions and names, reorganized issues, 
added impacts, and recorded facets describing the different ways each issue can manifest. 
No issue assignments were accepted solely on the basis of majority voting. 
The final classifications therefore reflect the outcome of the discussions rather than the initial card sorting results.
Although the detailed reasoning behind individual discussion decisions was not documented, 
the grouping of logging issues into smells was primarily guided by their \emph{primary manifestation}, 
that is, the aspect of logging in which the problem becomes apparent. 
For example, issues primarily manifested in static log messages were grouped under Misleading Messages (LS5), 
whereas those whose primary manifestation lies in the dynamic values included in log entries were grouped under 
Deceptive Variable (LS4). Different forms of problems within the same category were represented as facets of 
the corresponding smell rather than as separate smells.

As the initial data extraction and taxonomy construction were completed by Author~1, we performed an additional validation step to assess the robustness of the resulting taxonomy. Author~2 independently extracted quotations related to logging issues from the included papers. These quotations were then compared with those identified by Author~1 to determine whether any quotations had been missed during the original extraction.
For each additionally identified quotation, Author~2 independently identified the corresponding logging issue without consulting the issue list produced by Author~1. The resulting issues were subsequently compared against Author~1's issue list to determine whether they had already been identified. For issues that were not present in the original list, we assessed whether they could be represented using the existing taxonomy or whether modifications to the taxonomy were required.

While this validation identified several additional issue descriptions and resulted in refinements 
and clarifications of issue descriptions, facet definitions, naming, and examples, all additionally 
identified issues could be mapped to the existing taxonomy. Consequently, the validation did not 
(a) lead to changes in the proposed structure of the taxonomy, (b) introduce new log smells, causes, or consequences, or (c) alter the categorization of the identified issues.

The outcome of the analysis was a set of 18 high-level logging issues
identified in the surveyed literature and categorized as log smells,
their causes, or their direct consequences. Additional issues not
related to log smells were grouped separately and are included in the
replication package. In the latter,  we include the full analysis results for Steps~1
and~2, including the identified issues and
the individual classifications assigned by each author in the open and
closed card sorts. While the outcomes of the consensus discussions are
reflected in the final taxonomy, we did not systematically record the
discussion process or the detailed rationale behind each individual
decision.

\medskip

\noindent\textbf{RQ2. Analyzing tool support for log smells.}
For RQ2, we identified tools reported in the surveyed literature that provide solutions for the log smells identified in RQ1. We focused on tools that suggest modifications or automatically repair issues. Each tool was analyzed with respect to the log smells defined in RQ1 to determine which smells were addressed by the identified tool support, and we mapped tools to the specific facets they address.

\subsection{Replicability}

To allow replication, verification, and extension of this study, the raw data and complete results of the study will be provided in a replication package. We plan to make it publicly available upon acceptance of the paper.

 \section{Results}
\label{sec:results}

\subsection{RQ1: Taxonomy of Log Smells}
\label{sec:RQ1_results}
An overview of the 18 identified logging issues, categorized as log smells or as their causes and consequences, is presented in Table~\ref{tab:taxonomy_overview}.
We describe each of the categories and their associated issues in more detail in the following subsections. 
To document the issues, we use the template presented in Figure~\ref{fig:example_log}. However, we have omitted or added elements for some issues based on their relevance.

\begin{table}
	\small
	\centering
	\caption{Overview of categorized logging issues.}
	\label{tab:taxonomy_overview}
	\begin{tabular}{p{5.4cm}p{4.2cm}p{4.5cm}}\toprule
		\textbf{Causes of Log Smells}	&	\textbf{Log Smells}	&	\textbf{Consequences of Log Smells}	\\	\midrule
		\begin{itemize}[leftmargin=0.7cm,  align=parleft]
			\raggedright
			\setlength\itemsep{0em}
			\item[CA1] Lack of General Logging Guidelines
			\item[CA2] Developers' Domain Knowledge and Experience
			\item[CA3] Logging Library Usage
			\item[CA4] Logging Integration
			\item[CA5] Insufficient Logging Code Maintenance
		\end{itemize} &
		\begin{itemize}[leftmargin=0.7cm, align=parleft]
			\setlength\itemsep{0em}
			\item[LS1] Format Drift
			\item[LS2] Undercover Identifier
			\item[LS3] Shifting Logging Levels
			\item[LS4] Deceptive Variable
\item[LS5] Misleading Messages
			\item[LS6] Logging Blackout
			\item[LS7] Landfill Logs
			\item[LS8] Sleeping Guards
			\item[LS9] Logging Code Debt
		\end{itemize} & 
		\begin{itemize}[leftmargin=0.7cm]
			\setlength\itemsep{0em}
			\raggedright
			\item[CO1] Information Leak
			\item[CO2] Temporal Inconsistency
			\item[CO3] Effect on System Performance
			\item[CO4] Accidental Side-effects of Logging Code
		\end{itemize} \\ \bottomrule
	\end{tabular}
\end{table}

\begin{figure}
	\caption{Template used for presenting the logging issues.}
	\begin{tcolorbox}[colback=white, boxrule=0.2mm]
		\noindent\textbf{(Template) Issue ID: Name of the issue}
		
		\scope{\\[0.1em]The areas of logging affected by the smell. 
		Possible values are Code, Entry, File, Parsing, and Downstream task.}
		\vspace{0.2em}
		
		\descri \vspace{0.3em}General description of the smell, 
		including where the issue primarily manifests and why it is problematic.
		\vspace{0.2em}
		
		\facets 
		\vspace{0.3em}Describes the different forms of the smell (facets) and their implications. For example:
		\begin{smellfacets}
			\facet{Missing}{The severity is not specified.}
			\impact{Relevant events will not be detected by severity based filtering.}
			\facet{Inconsistent}{Similar events have different logging levels in different parts of the code.}
			\impact{Produces unpredictable filtering and reduces confidence in the meaning of logging levels.}
		\end{smellfacets}
		
		\references{References to papers mentioning the issue.}
		
		\vspace{0.2em}
		\example  \vspace{0.3em}Provides an illustrative example of log entries or logging code affected by the issue, with problematic parts shown in gray. When applicable, we modify the example provided below. We also provide references to real-world examples from existing literature.
		
		\begin{Verbatim}[fontsize=\footnotesize, commandchars=\\\{\}, breaklines=true, xleftmargin=-1.2cm]
			1 2024-07-26 12:39:19Z INFO [database] Connecting to database
			2 2024-07-26 12:41:46Z WARN [database] Connection slow, retrying
			3 2024-07-26 12:42:16Z ERROR [database] Failed to establish connection: Connection timeout. Abort proc.
			...
			n 2024-07-26 16:34:12Z INFO [database2] Connecting to database
		\end{Verbatim}
		
	\end{tcolorbox}
	\label{fig:example_log}
\end{figure}

\begin{figure}
	\centering
	\small
	\tikzset{
	level1/.style={
		minimum width=3.9cm,
		fill=black!3,
	},
	level2/.style={
		minimum width=3.5cm,
		fill=black!6,
	},
	smell/.style={
	minimum width=4.1cm,
	fill=black!9
}
}

\begin{forest}                      
	forked edges,                        
	for tree={grow=0, 
		draw,
		align=left,
} 
	[Log Smell    
		[Logging Code (white box) , level1
			[Invisible to the Logs, level2
				[Logging Code Debt, smell]] 
			[Transparent to the Logs, level2
				[Sleeping Guards, smell]]  
		]
		[Log Artifact (log file), level1
			[Log Files,level2
				[Landfill Logs, smell]
				[Logging Blackout, smell]]
			[Log Entry Content, level2
				[Deceptive Variable, smell]
				[Misleading Messages, smell]
				[Undercover Identifier, smell]
				[Shifting Logging Levels, smell]
				[Format Drift, smell]]
		] 
	] 
\end{forest}

 	\caption{Taxonomy of log smells.}
	\label{fig:smell_taxonomy}
\end{figure}

\subsubsection{Log Smells}
\label{sec:log_smells}

This section introduces the nine logging issues that we classify as log smells and presents the resulting taxonomy.
Figure~\ref{fig:smell_taxonomy} shows the taxonomy, which classifies log smells based on the aspect of logging in which 
the issues primarily manifest.

The smells were formed by grouping issues according to the primary aspect of logging in which they manifest. 
Issues with the same primary manifestation were grouped 
into a single log smell, while their different forms 
were represented as facets rather than separate smells.
The first level differentiates smells based on whether they primarily manifest in the \emph{log artifact} 
(log file) or the \emph{logging code}. Since the logs are produced by the code, all problems can be associated with it to some extent. Nevertheless, we considered 
that a smell is related to the code if it manifests in the logic or cleanliness of the code. 
For example, including an unnecessary log line was not regarded as a coding issue, 
whereas omitting an if-clause related to logging was. 
Smells that manifest in the 
logging code are further divided into those that are
\emph{transparent to the logs}, meaning that their effects
are also visible in the generated log artifact, and those
that are \emph{invisible to the logs}, meaning that the
problem remains confined to the logging code.
The smells that manifest in the log artifact are divided according to whether they 
concern the \emph{log files} or are specific to the \emph{log entry content}. 

When applying the taxonomy, users should consider the level of abstraction at which they wish to classify issues. The taxonomy was designed around the primary manifestation of a logging issue, but it can also be applied to identify more specific manifestations represented by individual facets. A single issue may involve multiple logging problems and therefore be associated with multiple log smells, particularly when considering facet-level manifestations. When determining the primary manifestation, we found it useful to consider the intended purpose of the logging change, as reflected in the issue description and commit message, together with the code change itself. If an issue does not fit an existing category, users should determine whether it represents a new log smell, a cause, a consequence, or another logging-related issue before extending the taxonomy.

We illustrate the log smells in practice by showcasing a smelly log or logging code for each type. 
When possible, we modified the log shown in Figure~\ref{fig:example_log} to demonstrate the corresponding
smell. 
For some smells, the log was not applicable, so a different example is used in those cases. 
In the modified examples, \textvisiblespace ~denotes a missing value or log entry, 
while $\hookrightarrow$ denotes a line break.

\setlength{\parindent}{0pt}  
\setlength{\parskip}{0.4em}

\smelltitle{LS1}{Format Drift}

	\scope{Entry, File, Parsing}
	
	\descri 
	Issues in the format of log entries. A well-defined format ensures that log entries follow a consistent structure, enabling effective log search and making logs more machine-friendly. Format-related issues increase the effort required to use information in log files, especially when information must be interpreted or searched across multiple files.
		
	\facets
	
	\begin{smellfacets}
			\facet{Incomplete}{The format specification is missing fields that are relevant for the use case of the logs.}
				\impact{Important contextual information may be unavailable.}
	
			\facet{Ambiguous}{The format is ambiguously specified.}
				\impact{Allows different interpretations and reduces formatting consistency.}
	
			\facet{Not following a format} { A logging format has been specified, but some log entries do not conform to it.}
				\impact{Reduces the reliability of automated processing and parsing.}
	
			\facet{Multiple formats}{Different components of the software use different formats for logging. }
				\impact{Makes it more difficult to aggregate, compare, and search log entries across components.}
	\end{smellfacets}

	\references{\cite{batoun_literature_2024}, \cite{bijvank_software_nodate}, \cite{chen_survey_2022}, \cite{ he_empirical_2022}, \cite{ Lu2018CloudRaid}, \cite{ marron_log_2018}, \cite{ patel_sense_2022}, \cite{ Qi2017ACloud}, \cite{ rong_how_2023}, \cite{Sedki2025Developing}, \cite{ yang_interview_2021}}
	
	\example
	The following two log entries originate from different systems within the same software project; however, the systems use different formats for logging the same event. Additionally, neither of the entries presents all the necessary
	information shown in the log example provided in Figure~\ref{fig:example_log}: line~1 lacks the verbosity level, while line~8 has an incomplete timestamp. These issues would create difficulties in filtering and matching the information. 
	
\begin{Verbatim}[xleftmargin=-1.2cm, fontsize=\footnotesize, commandchars=\\\{\}, breaklines=true]
	1 [16:00:34.276Z 08/06] [database] \colorbox{gr}{\textvisiblespace} Connecting to database DB1
	...
	8 INFO \colorbox{gr}{Jun 08 24} sys2: DB2 connecting to database
\end{Verbatim}
	
	A real-life example of a formatting issue is presented in~\cite[p.33 (LF12)]{patel_sense_2022}.

\smelltitle{LS2}{Undercover Identifier}

	\scope{Entry, File, Parsing, Downstream Task}
	
	\descri 
	Issues in the identifier of a log entry. Identifiers are used to distinguish the source or execution context of log entries, such as components and threads. Correct identifiers enable log entries to be linked together and associated with the entities that produced them, making log analysis and troubleshooting more effective.
		
	\facets
	
	\begin{smellfacets}
		\facet{Missing}{The log entry is missing the identifier.}
		\impact{The entry cannot be reliably associated with its originating component or execution context, making analysis difficult or impossible. This is especially critical in distributed and concurrent systems.}
		\facet{Wrong identifier}{The identifier in the log entry is incorrect.}
		\impact{The log entry is associated with the wrong component or execution context, potentially leading to incorrect conclusions during analysis.}			
	\end{smellfacets}

	\references{ \cite{ bogatinovski_qulog_2022}, \cite{ li_qualitative_2021}, \cite{ Lu2018CloudRaid}, \cite{ Qi2017ACloud}, \cite{xu_detecting_2009}, \cite{ zhao_studying_2023}}

	\example 
		Similar to the log example in Figure~\ref{fig:example_log}, line~1 correctly indicates that a connection is being established. In contrast, line~2 is incorrectly labeled with the server component, even though it should report the slow database connection. This can be misleading, as a slow server connection is also possible. Line~3 lacks an identifier, making it unclear where the error occurred.
	
\begin{Verbatim}[xleftmargin=-1.2cm, fontsize=\footnotesize, commandchars=\\\{\}, breaklines=true]
	1 2024-07-26 12:39:19Z INFO [database] Connecting to database
	2 2024-07-26 12:41:46Z WARN \colorbox{gr}{[server]} Connection slow, retrying
	3 2024-07-26 12:42:16Z ERROR \colorbox{gr}{\textvisiblespace} Failed to establish connection: Connection timeout. Abort proc.
\end{Verbatim}
	
	A real-life example of a missing identifier is presented in \cite[Listing 1]{zhao_studying_2023}.

\smelltitle{LS3}{Shifting Logging Levels}

	\scope{Entry, Downstream Task}
	
	\descri 
	Issues in logging level. The level indicates the importance of a log entry and supports log categorization, filtering, and prioritization. However, assigning the appropriate level is often difficult. Issues in the level reduce the effectiveness of filtering and analysis, as important events may be overlooked or obscured by noise.
	
	\facets

	\begin{smellfacets}
	\facet{Missing}{The severity is not specified.}
		\impact{Relevant events will not be detected by severity-based filtering.}
	\facet{Incorrect}{The assigned level does not reflect the event (e.g., marking an expected event as an error).}
		\impact{Important events may be overlooked, while irrelevant events may receive unnecessary attention.}
	\facet{Inconsistent}{Similar events have different logging levels in different parts of the code.}
		\impact{Produces unpredictable filtering and reduces confidence in the meaning of logging levels.}
	\end{smellfacets}
	
	\references{ \cite{anu_approach_2019}, \cite{batoun_literature_2024}, \cite{ bogatinovski_qulog_2022}, \cite{ chen_characterizing_2017}, \cite{ chen_extracting_2019}, \cite{chen_survey_2022}, \cite{ gholamian_logging_2020}, \cite{ hassani_studying_2018}, \cite{ Kulkarni2024}, \cite{ li_qualitative_2021}, \cite{ li_towards_2017}, \cite{ li_deeplv_2021}, \cite{ Liu2022}, \cite{ Oliner2012}, \cite{ patel_sense_2022}, \cite{ rong_how_2023}, \cite{ Wang2025}, \cite{ xu_detecting_2009}, \cite{ Yu2023LogReducer}, \cite{ yuan_zing_2012}, \cite{ zhao_game_2017}, \cite{ zhaoxue_survey_2021}}
	
	\example 
		In the log example based on Figure~\ref{fig:example_log}, line~2 incorrectly labels a slow database connection as an ERROR rather than a WARNING. Line~3 then reports the timeout without any log level, although it should be marked as an ERROR. Finally, line~81 logs the same ``connecting to database'' message as in line~1 but assigns it a different level, creating an inconsistency.
\begin{Verbatim}[xleftmargin=-1.2cm, fontsize=\footnotesize, commandchars=\\\{\}, breaklines=true]
	1 2024-07-26 12:39:19Z INFO [database] Connecting to database
	2 2024-07-26 12:41:46Z \colorbox{gr}{ERROR} [database] Connection slow, retrying
	3 2024-07-26 12:42:16Z \colorbox{gr}{\textvisiblespace} [database] Failed to establish connection: Connection timeout. Abort proc.
	...
	81 2024-07-26 16:34:12Z \colorbox{gr}{DEBUG} [database2] Connecting to database
\end{Verbatim}
	
	Real-life examples are provided, among others, in \cite[Fig 4]{chen_characterizing_2017}, \cite[Tab 15]{patel_sense_2022}, \cite[p. 3262]{hassani_studying_2018}, \cite[Fig 1b]{bogatinovski_qulog_2022}, and \cite[p. 1842]{li_towards_2017}.

\smelltitle{LS4}{Deceptive Variable}

	\scope{Entry, File, Parsing, Downstream Task}
	\descri 
	Issues in the variables included in log entries. These variables contain dynamic runtime information, such as parameter values and execution results. Correct variables provide the context needed to understand events, correlate related log entries, and support troubleshooting and analysis.

		\facets

	\begin{smellfacets}
		\facet{Missing}{The appropriate variable is omitted.}
		\impact{Important context may be unavailable, making events harder to interpret and analyze.}
		
		\facet{Wrong}{An incorrect variable is logged.}
		\impact{May misrepresent the state of the system, lead to incorrect conclusions during analysis, or expose sensitive information.}
		
		\facet{Malformed output}{The logged variable does not have a meaningful or well-defined textual representation.}
		\impact{Might produce a long and confusing output which is difficult to interpret and process.}
		
		\facet{Varying granularity}{Comparable events are logged using different levels of detail or different sets of variables.}
		\impact{May result in information loss and make similar code appear to behave differently.}
	\end{smellfacets}

	\references{ 
		\cite{batoun_literature_2024}, \cite{bijvank_software_nodate}, \cite{ chen_characterizing_2017}, \cite{ chen_extracting_2019}, \cite{ patel_sense_2022}, \cite{ rong_how_2023}, \cite{Wang2025}, \cite{ yuan_zing_2012}, \cite{Zhong2025LogUpdater}}
	
	\example
		In the example based on Listing~\ref{lst:logging_code_example}, line~5 logs the variable \verb|conn| instead of \verb|e|. Because \verb|conn| is an object, not an error description, this may produce malformed or irrelevant output if the object lacks a string representation. Additionally, without logging the correct variable, the failed connection cannot be linked to the relevant database, which is critical for debugging.
	
\begin{Verbatim}[xleftmargin=-1.2cm, fontsize=\footnotesize, commandchars=\\\{\}, breaklines=true, numbers=left, xleftmargin=5mm]
try:
    logger.info(f'Connecting to database')
    conn = sqlite3.connect(db_file)
except sqlite3.Error as e:
    logger.error(f'Failed to establish connection: \colorbox{gr}{\{conn\}}. Abort proc.')
\end{Verbatim}
	
	An example of the \emph{varying granularity} facet relates to the accuracy with which a variable is logged. Consider the exchange‑rate example: the code logs the rate when it exceeds a predefined threshold with six decimals, but the logged value is printed with only three.
	
\begin{Verbatim}[xleftmargin=-1.2cm, fontsize=\footnotesize, commandchars=\\\{\}, breaklines=true, numbers=left, xleftmargin=5mm]
if exchange_rate >= \colorbox{gr}{1.555555}:
    logging.warning(f"High exchange rate for {currency_pair}: \{exchange_rate:.\colorbox{gr}{3}f\}")
\end{Verbatim}
	
	Real-life examples are presented, for example, in \cite[Fig 5]{yuan_zing_2012}, \cite[Fig 4]{chen_characterizing_2017}, and \cite[Tab 14, example 7]{patel_sense_2022}.

\smelltitle{LS5}{Misleading Messages}
        \label{smell:ls5:mm}

	\scope{Entry, File, Downstream Task}
	\descri 
	Issues in the static message of a log entry. The message describes the event being logged and provides human-readable context. Clear and accurate messages help practitioners understand events and interpret logs correctly, while issues in the message can hinder comprehension and analysis.

	\facets

	\begin{smellfacets}
		\facet{Missing}{The static message is completely missing.}
		\impact{Important context is unavailable, making comprehension and analysis more difficult.}  
		
		\facet{Wrong}{The static part of the log message is wrong.} 
		\impact{Makes the log confusing and might cause wrong analysis results.} 
		
		\facet{Imprecise} {The message is ambiguous or omits crucial information. }
		\impact{May cause ambiguity and make events more difficult to interpret.}
		
		\facet{Language issues} {Issues in the language of the message such as typos, grammatical errors (such as incorrect verb tense), unclear wording, or inaccurate translations. }
		\impact{Reduces readability and increases the effort required to interpret log messages.}
		
		\facet{Inconsistent}{ Messages associated with similar events differ even though they should be consistent. This is an issue across multiple entries, not a single one. }
		\impact{Reduces message predictability and makes related events more difficult to identify and compare.}
		
		\facet{Duplicated}{ Identical messages appearing in different log templates. Issue across multiple entries, not a single one. }
		\impact{Makes locating the corresponding position in source code more difficult.}
	\end{smellfacets}

\references{ 
	\cite{batoun_literature_2024}, \cite{bijvank_software_nodate}, \cite{Bogatinovski2023Auto}, \cite{ bogatinovski_qulog_2022}, \cite{ chen_extracting_2019}, \cite{ chen_survey_2022}, \cite{ ding_temporal_2023}, \cite{ gu_logging_2023}, \cite{ he_characterizing_2018-1}, \cite{ hassani_studying_2018}, \cite{ he_empirical_2022}, \cite{ li_qualitative_2021}, \cite{ li_towards_2017}, \cite{ li_studying_2021}, \cite{ li_studying_2022},  \cite{ oliveira_automatically_2020}, \cite{ lu_roundtable_2015}, \cite{ Oliner2012}, \cite{ patel_sense_2022}, \cite{ rong_how_2023}, \cite{Sedki2025Developing}, \cite{ shen_improving_nodate}, \cite{Wang2025}, \cite{ yuan_zing_2012}, \cite{ zhao_studying_2023}, \cite{ Zhong2025LogUpdater} }

	\example
	In the log example based on Figure~\ref{fig:example_log}, line~1 uses the wrong verb tense, incorrectly suggesting the connection is already established. It is also inconsistent with line~8, even though both describe the same event. Line~2 has unclear wording, where ``slow'' would be more precise. Finally, line~3 omits the message about the failed connection.
	
	\begin{Verbatim}[xleftmargin=-1.2cm, fontsize=\footnotesize, commandchars=\\\{\}, breaklines=true]
		1 2024-07-29 15:09:51Z INFO [database] Connect\colorbox{gr}{ed} to database
		2 2024-07-29 15:11:36Z WARN [database] Connection \colorbox{gr}{stagnant}, retrying
		3 2024-07-29 15:12:03Z [database] ERROR \colorbox{gr}{\textvisiblespace}
		...
		8 2024-07-26 16:34:12Z INFO [database2] Connecting to database
	\end{Verbatim}
	
	Real-life examples are provided, for example, in \cite[Fig 8]{chen_extracting_2019}, \cite[Fig 9]{yuan_zing_2012}, and \cite[Tab 13]{patel_sense_2022}.

\smelltitle{LS6}{Logging Blackout}

	\scope{Code, Entry, File,  Downstream Task}
	\descri 
	Issues related to missing logging information. This smell occurs when information that would be useful for understanding system behaviour is absent from the logs. As a result, important information may be unavailable for monitoring, troubleshooting, and analysis.
	
	\facets

	\begin{smellfacets}
		\facet{Overlooked}{A relevant logging statement is missing from the code.}
		\impact{Reduces the usefulness of the logs, as important runtime information is not recorded and some issues may go unnoticed. For example, missing entries break the continuity of execution flow, making it difficult to reconstruct system behavior or recognize expected structures.}
		\facet{Hidden}{ Expected log entries are not generated or emitted because of logging configuration issues (e.g., the logging level is set too high).
}
		\impact{Hides important runtime information, reducing log visibility and making issue detection and diagnosis more difficult.
} 
	\end{smellfacets}
	
	\references{  
		\cite{batoun_literature_2024}, \cite{chen_extracting_2019}, \cite{chen_survey_2022}, \cite{ Chen2020Understanding}, \cite{ fu_where_2014}, \cite{ hassani_studying_2018}, \cite{ he_empirical_2022}, \cite{ Kulkarni2024}, \cite{ li_qualitative_2021}, \cite{ li_towards_2017}, \cite{ li_guiding_2020}, \cite{ oliveira_automatically_2020}, \cite{ Lu2018CloudRaid}, \cite{ lu_roundtable_2015}, \cite{Mace2018Pivot}, \cite{ patel_sense_2022}, \cite{ Qi2017ACloud}, \cite{ rong_how_2023}, \cite{ shen_improving_nodate}, \cite{Wang2025}, \cite{ yang_interview_2021}, \cite{ yuan_be_nodate}, \cite{ Zamfirov2024}, \cite{ zhaoxue_survey_2021}, \cite{ zhu_learning_2015} }
	
	\example
		Compared to the log example provided in Figure~\ref{fig:example_log}, the example below is missing a log entry. It does not indicate whether a connection to the database was established. This can complicate debugging, as there is no record of the connection failing due to a timeout.
\begin{Verbatim}[xleftmargin=-1.2cm, fontsize=\footnotesize, commandchars=\\\{\}, breaklines=true]
	1 2024-07-29 15:09:51Z INFO [database] Connecting to database
	2 2024-07-29 15:11:36Z WARN [database] Connection slow, retrying
	\textbf{\colorbox{gr}{\textvisiblespace}}
	...
	8 2024-07-26 16:34:12Z INFO [database2] Connecting to database
\end{Verbatim}
	
	Real-life examples are provided in \cite[Sec 4.3.1]{patel_sense_2022} and \cite[Fig 8 (AMI)]{chen_extracting_2019}.

\smelltitle{LS7}{Landfill Logs}

	\scope{Entry, File, Downstream Task}

	\descri 
	 Log entries or log files overloaded with low-value information. This smell occurs when the amount of information recorded exceeds what is useful for the intended purpose of the log. It can reduce the usefulness of logs by obscuring relevant information and increasing the effort required to analyse them. The smell is context dependent, as whether information is considered useful or excessive depends on the task for which the logs are used.

	\facets
	
	\begin{smellfacets}
		\facet{Useless}{The log entry does not provide information that is valuable for the intended task.}
		\impact{Obscures relevant information and increases analysis effort.}
		\facet{Redundant}{Information is recorded repeatedly across log entries, for example due to repeated logging within loops.}
		\impact{Inflates log volume without providing additional value.}
		\facet{Too detailed}{The log entry contains more detail than required for the intended purpose, for example because information is logged without appropriate aggregation.}
		\impact{Increases storage costs and reduces log comprehensibility.}
		\facet{Stack trace overuse} {The stack trace is logged without a justified need for it. }
		\impact{Increases log volume and makes relevant information harder to locate.}
				
	\end{smellfacets}
	
	\references{ 
		\cite{batoun_literature_2024}, \cite{Bogatinovski2023Auto}, \cite{ chen_extracting_2019}, \cite{chen_survey_2022}, \cite{ding_log2_nodate}, \cite{ fu_where_2014}, \cite{ gu_logging_2023}, \cite{ hassani_studying_2018}, \cite{ he_empirical_2022}, \cite{ Jia2022Augmenting}, \cite{ Kulkarni2024}, \cite{ li_qualitative_2021}, \cite{ li_guiding_2020}, \cite{ li_studying_2022}, \cite{ lu_roundtable_2015}, \cite{Mace2018Pivot}, \cite{ marron_log_2018}, \cite{ patel_sense_2022}, \cite{ rong_how_2023}, \cite{ yang_interview_2021}, \cite{ zeng_studying_2019}, \cite{zhou_mobilogleak_2020}, \cite{ zhu_learning_2015} }
	
	\example
	Compared to the log example in Figure~\ref{fig:example_log}, the example below shows that the slow connection in line~3470 and the eventual failure in line~7392 appear thousands of lines apart due to excessive logging. In addition, line~3471 is unnecessary, while line~3472 is relevant but likely too detailed for most use cases.
	
\begin{Verbatim}[xleftmargin=-1.2cm, fontsize=\footnotesize, commandchars=\\\{\}, breaklines=true]
	1 2024-07-26 12:39:19Z INFO [database] Connecting to database
	...
	\colorbox{gr}{3470} 2024-07-26 12:41:46Z WARN [database] Connection slow, retrying
	\colorbox{gr}{3471 2024-07-26 12:40:46Z DEBUG: The quick brown fox jumps over the lazy dog.} 
	\colorbox{gr}{3472 2024-07-26 12:41:46Z INFO Server started on port 8080 at 2024-07-26 12:41:43 | Environment: Prod |} \colorbox{gr}{Additional Notes:  Backup Schedule: Daily, 02:00 UTC, Log Rotation: Weekly, Sundays at 00:00 UTC}
	...
	\colorbox{gr}{7392} 2024-07-26 12:42:16Z ERROR [database] Failed to establish connection: Connection timeout. Abort proc.
\end{Verbatim}
	
	Real-life examples are provided in \cite[p. 3265]{hassani_studying_2018}, \cite[Fig 8h]{chen_extracting_2019}, and \cite[Tab 15 example 9]{patel_sense_2022}.

\smelltitle{LS8}{Sleeping Guards}

	\scope{Code}
	
	\descri 
	Issues in logging guards. Logging guards control whether logging code executes, typically based on the configured logging level. Correct guards avoid unnecessary computation when log entries will not be recorded and ensure that logging behaviour remains consistent with the intended configuration.
	
	\facets

	\begin{smellfacets}
		\facet{Missing}{The logging guards are missing, which allows logging code to run even when no entry will be recorded.}
			\impact{Wastes computational resources and may degrade system performance.}
		\facet{Incorrect}{The conditions used in the logging guards do not match their intended behaviour, such as requiring a different logging level than the one used in the corresponding statement.}
			\impact{Can leave variables undefined or outdated, potentially causing crashes or producing misleading log entries.}
	\end{smellfacets}
	
	\references{ 
		\cite{ batoun_literature_2024}, \cite{ chen_extracting_2019}, \cite{ li_qualitative_2021}, \cite{ Yu2023LogReducer}, \cite{ zhi_towards_2022}  }
	
	\example
		This smell is illustrated with code that retrieves an HTML page and generates a report intended to be logged only when the program’s logging verbosity is set to DEBUG. Without logging guards, the report is generated every time the code runs, regardless of the current verbosity. Producing the report may require considerable resources, depending on the size of the web page.
	
\begin{Verbatim}[ fontsize=\small, commandchars=\\\{\}, numbers=left, xleftmargin=5mm]
logging.debug("Starting webpage processing")
html = fetch_webpage(url)
titles = parse_webpage(html)
\colorbox{gr}{<missing logging guard>}
report = generate_debug_report(titles)
report_str = "Report Summary: \{report['summary']\}, Titles: ..."
logging.debug("Generated report: \{report_str\}")    
\end{Verbatim}
	A real-life example is provided in \cite[Fig 9 (ALG)]{chen_extracting_2019}.

\smelltitle{LS9}{Logging Code Debt (Logging Code Smells)}

		\scope{Code}
		
		\descri 
			Issues related to the code used to create the log. This includes logging statements themselves as well as supporting code, such as condition checks, data aggregation, formatting, and variable conversions. Unlike other log smells, this smell primarily affects the maintainability and comprehensibility of the code that produces logs rather than the produced log entries themselves. It is therefore a \emph{meta log smell}, as it affects the code that produces logs rather than the content of the logs themselves.
	
	\facets

	\begin{smellfacets}
		\facet{Code smells}{Traditional software code smells affecting logging related code, such as duplicated code, long methods, temporary variables, tangled logging and feature code, or unnecessary complexity.}
		\impact{Reduce maintainability and increase the effort required to modify or evolve logging code.}
		\facet{Poor comprehensibility}{The intent or behaviour of logging code is difficult to understand due to poor naming, confusing structure, or insufficient documentation.}
		\impact{Increases the likelihood of maintenance mistakes and makes logging behaviour harder to understand.}
		\end{smellfacets}

	\references{ 
		\cite{ batoun_literature_2024}, \cite{ chen_characterizing_2017}, \cite{ chen_extracting_2019}, \cite{ chen_survey_2022}, \cite{ gholamian_logging_2020}, \cite{ li_qualitative_2021}, \cite{ zhao_studying_2023}, \cite{ zhi_towards_2022}, \cite{ zhu_learning_2015} }
		
	\example		
		The function below logs a payment. The variables on lines~2 and~3 are poorly named and used only for logging, making them ``Temporary Variables''. Variable \verb*|a| is used on line~6 but not on line~10, where the code is duplicated. The conversion on line~3 is unnecessary, and the \verb|if| clause on line~7 adds needless complexity.
	
\begin{Verbatim}[xleftmargin=-1.55cm, fontsize=\small, commandchars=\\\{\}, numbers=left, xleftmargin=5mm]
def log_payment(payment):
    \colorbox{gr}{a} = str(payment['amount']) + " USD"
    \colorbox{gr}{b} = str(payment['customer'])\colorbox{gr}{.upper()}
	
    if payment.get('status') == 'SUCCESS':
        msg = "Payment successful for " + b + " of " + a
        \colorbox{gr}{if len(msg) > 0:}  
            logging.info(msg)
    else:
        msg = "Payment failed for " + b + " of " + \colorbox{gr}{str(payment['amount']) + " USD"}
        logging.info(msg)
\end{Verbatim}
	
	Real-life examples are provided in \cite[Fig 4]{chen_characterizing_2017} and  \cite[Tab 14 examples 5 and 6]{patel_sense_2022}.

\subsubsection{Causes of log smells}
\label{sec:causes_of_ls}

The section presents in detail the five logging issues classified as causes of log smells. \\  

\smelltitle{CA1}{Lack of General Logging Guidelines}

	\descri 
		Log smells commonly arise when developers do not share a consistent understanding of logging practices. Developers or teams may adopt different conventions, creating the need for clear, project‑level specifications. 
		
		\facets
		\begin{smellfacets}
			\facet{Missing}{Lack of specifications forces developers to rely on individual preferences and experience when making logging decisions, increasing variation in logging practices.}
			\facet{Incomplete}{Some aspects of logging are standardized while others are left unspecified, allowing developers to interpret the missing guidance differently and creating unnecessary inconsistency.}
			\facet{Outdated}{Guidelines no longer reflect the current state of the system or development practices, reducing their usefulness, causing confusion, and requiring additional effort to align logging with project needs.}

		\end{smellfacets}

	\references{
		\cite{ anu_approach_2019}, \cite{ Bogatinovski2023Auto}, \cite{ bogatinovski_qulog_2022}, \cite{ gholamian_logging_2020}, \cite{ he_characterizing_2018-1}, \cite{ li_guiding_2020}, \cite{ Li2024}, \cite{ li_studying_2022}, \cite{Mace2018Pivot}, \cite{ Oliner2012}, \cite{ patel_sense_2022}, \cite{ rong_how_2023}, \cite{ yuan_zing_2012}, \cite{ Zamfirov2024}, \cite{zhou_mobilogleak_2020}, \cite{ zhu_learning_2015}}

\smelltitle{CA2}{Developers' Domain Knowledge and Experience}

	\descri  		
		Logging decisions are made by the developers implementing the logging code, yet these decisions depend heavily on their experience and familiarity with the technologies, development practices, and the system under development. As a result, logging practices may differ even when common guidelines exist. Developers who lack \textit{domain knowledge} or are not \textit{familiar with the entire project} may have difficulty making appropriate logging decisions, particularly in multi‑component systems where interpreting logs is more complex.

	\implications
		Creates unnecessary differences in logging practices and logs. 
	
	\references{ 
		\cite{ anu_approach_2019}, \cite{ Bogatinovski2023Auto}, \cite{ bogatinovski_qulog_2022}, \cite{ gu_logging_2023}, \cite{ he_empirical_2022}, \cite{ li_studying_2022}, \cite{ oliveira_automatically_2020}, \cite{Mace2018Pivot}, \cite{ patel_sense_2022}, \cite{ rong_how_2023}, \cite{ yang_interview_2021}, \cite{ Zamfirov2024}, \cite{ zhao_game_2017}, \cite{ zhaoxue_survey_2021}, \cite{ zhu_learning_2015}}

\smelltitle{CA3}{Logging Library Usage}

	\descri
	  Software projects often rely on logging libraries and frameworks to provide systematic logging. However, challenges may arise from the selection, coexistence, configuration, and integration of these libraries. Such library-related issues can make logging harder to standardize, maintain, and understand across a project.

	\facets
	\begin{smellfacets}
		
		\facet{Several libraries}{The project uses multiple logging libraries or frameworks, which may lead to inconsistencies in supported features, log formats, configuration options, and dependency requirements.}

		\facet{Wrong or inconsistent configuration}{Logging libraries or frameworks are incorrectly configured or inconsistently configured across components, environments, or dependencies. This might lead to missing log entries, excessive log verbosity, dependency conflicts, permission-related issues, and inconsistent log output.}

	\end{smellfacets}
	
	\references{ 
		\cite{ batoun_literature_2024}, \cite{ chen_extracting_2019}, \cite{ chen_survey_2022}, \cite{ Chen2020Understanding}, \cite{hassani_studying_2018} }

\smelltitle{CA4}{Logging Integration}

	\descri
	     Modern software systems often consist of multiple
             components, depend on several technologies and execution
             contexts, and are developed by different teams. Integrating consistent logging across these parts can be challenging, as they may rely on different logging libraries, formats, configurations, and development practices.  This can make logging difficult to standardize, correlate, and analyze across the project.

	\facets
	\begin{smellfacets}
		
		\facet{ Structural}{Logs produced by
                    independently developed components or different
                    technologies (e.g., logging frameworks) may use varying logging practices, formats, structures, and storage locations. These differences can reduce consistency and interoperability, complicating the aggregation, comparison, and analysis of logs across the system.}
		\facet{Runtime}{Logs generated by
                    distributed, concurrent, or multi-process
                    environments can make log collection, ordering,
                    correlation, and analysis more complex. For
                    example, such a scenario may obscure relationships between events and execution flows, making troubleshooting and system behavior reconstruction more challenging.}
			\end{smellfacets}
	\references{ 
		\cite{ batoun_literature_2024}, \cite{ bijvank_software_nodate}, \cite{ Bogatinovski2023Auto}, \cite{ he_characterizing_2018-1}, \cite{ he_empirical_2022}, \cite{ Jia2022Augmenting}, \cite{Mace2018Pivot}, \cite{ marron_log_2018}, \cite{ Oliner2012}, \cite{ Qi2017ACloud}, \cite{ rong_how_2023}, \cite{xu_detecting_2009}, \cite{ yang_interview_2021}, \cite{ zeng_studying_2019}}

\smelltitle{CA5}{Insufficient Logging Code Maintenance}

	\descri 			
		Developers frequently update feature code, but keeping the corresponding logging code aligned can be difficult. Logging statements are often written or refined in an ad‑hoc manner, sometimes only after failures occur and logs are needed.
	
	\implications	
		Insufficient maintenance may eventually lead to logging code that becomes obsolete, insufficient, or misleading. Updates to logging code do not necessarily improve it, as changes may address only a specific use case while overlooking other aspects. Such changes can introduce errors, including typos, wrong variables, or reduced comprehensibility. Maintenance practices also vary across domains, which can influence how logging code evolves over time.
	
	\references{ 
		\cite{ Bogatinovski2023Auto}, \cite{ chen_extracting_2019}, \cite{ chen_survey_2022}, \cite{ fu_where_2014}, \cite{ gu_logging_2023}, \cite{he_characterizing_2018-1}, \cite{ li_qualitative_2021}, \cite{ li_towards_2017}, \cite{ li_guiding_2020}, \cite{ patel_sense_2022}, \cite{ Qi2017ACloud}, \cite{ rong_how_2023}, \cite{ shen_improving_nodate}, \cite{ yang_interview_2021}, \cite{ yuan_zing_2012}, \cite{ Zamfirov2024}, \cite{ zeng_studying_2019}, \cite{ zhu_learning_2015}}

\subsubsection{Consequences of log smells}
\label{sec:consequences_of_ls}

This section describes the consequences of log smells.
As mentioned earlier, we focused only on the direct consequences of log smells. 

\smelltitle{CO1}{Information Leak}

	\descri 
		The log files of a software system may be publicly accessible to its users. In such cases, including sensitive information in the logs can lead to an information leak. Such data may involve network, database, location, or user account details, as well as kernel pointers and cryptographic keys.
	
	\implications
		Leaked information may make the software more vulnerable to attacks, as the exposed data can be used against the system. Alternatively, if the leaked data contains personal or organizational information, it may be exploited directly or result in a privacy breach.
	
	\causes 
		``Shifting Logging Levels'' (LS3) may cause information leaks when log entries intended for internal debugging contain sensitive data but are assigned an incorrect logging level. A similar issue can occur with ``Sleeping Guards'' (LS8), where poorly defined guards allow sensitive information to be logged unintentionally. As part of ``Landfill Logs'' (LS7), overly detailed logs may also expose sensitive data by accident. Finally, reduced comprehensibility in the logging code, as described by ``Logging Code Debt'' (LS9), can make such leaks more likely.
	
	\references{ 
		\cite{ batoun_literature_2024}, \cite{ Bogatinovski2023Auto}, \cite{gu_logging_2023}, \cite{Kulkarni2024}, \cite{li_qualitative_2021}, \cite{li_towards_2017}, \cite{ patel_sense_2022}, \cite{ Wang2025}, \cite{ zeng_studying_2019}, \cite{ zhou_mobilogleak_2020}}
	
	\example  The example code connects to a database and logs all of the retrieved data. If access to the resulting log file is not restricted, it could leak sensitive user information.
\begin{Verbatim}[fontsize=\small, commandchars=\\\{\}, numbers=left, xleftmargin=5mm]
conn = sqlite3.connect(db_file)
cursor = conn.cursor()
cursor.execute("SELECT * FROM users WHERE id= ...)
user_data = cursor.fetchone()
\colorbox{gr}{logging.info("Fetched data: {user_data}")}
conn.close()
\end{Verbatim}
	
A real example is provided in example 13 of Table 17 in \cite{patel_sense_2022}.

\smelltitle{CO2}{Temporal Inconsistency}

	\descri 
		Temporal inconsistency arises when the temporal order of events in the code differs from the order inferred from the log file. For example, a log entry may indicate that a process has completed when it has only just started, or concurrency may cause log entries to appear in an order that does not reflect the actual sequence of actions. This issue is particularly problematic in a black‑box setting, where the source code is not available for verifying the correct order of events.
	
	\causes 	
		``Misleading Messages'' (LS5) can cause temporal inconsistency when language mistakes in a logging message---particularly verb tense---create a misleading impression of the code’s execution. Missing log entries, ``Logging Blackout'' (LS6), can also make it difficult to link related events and thus confuse the inferred order of actions. The same applies to ``Undercover identifier'' (LS2), where incorrect or missing identifiers hinder connecting entries that belong together. Finally, the reduced comprehensibility of the logging code associated with ``Logging Code Debt'' (LS9) can increase the likelihood of such inconsistencies.
	
	\implications
		Interpreting log files becomes more difficult, confusing, and laborious.
	
	\references{ 
		\cite{ding_temporal_2023}, \cite{ li_qualitative_2021}, \cite{ zhao_studying_2023}}
		
	\example 
		The example is based on the logging code shown in Listing~\ref{lst:logging_code_example} and the log file in Figure~\ref{fig:example_log}. The code at line~2 uses the wrong verb tense, even though it executes before the connection attempt. As a result, line~1 of the log file suggests the connection is already established. This misleads the reader when later entries report a slow connection and a failure to connect, even though the connection was never established and the system had already stopped trying.
	
	Logging code:
\begin{Verbatim}[fontsize=\footnotesize, commandchars=\\\{\}, breaklines=true, numbers=left, xleftmargin=5mm]
try:
    logger.info('Connect\colorbox{gr}{ed} to database')
    conn = sqlite3.connect(db_file)
except sqlite3.Error as e:
    logger.error('Fail\colorbox{gr}{ing} to connect to database: \{e\}. Abort proc.')
\end{Verbatim}
	
	Log file:
\begin{Verbatim}[xleftmargin=-1.2cm, fontsize=\footnotesize, commandchars=\\\{\}, breaklines=true]
	1 2024-07-29 15:09:51Z INFO [database] Connect\colorbox{gr}{ed} to database
	2 2024-07-29 15:11:36Z WARN [database] Connection slow, retrying
	3 2024-07-29 15:12:03Z [database] ERROR Fail\colorbox{gr}{ing} to connect to database: Connection timeout. Abort proc.
\end{Verbatim}
	
	Additional examples of temporal inconsistencies are presented in~\cite{ding_temporal_2023}.

\smelltitle{CO3}{Effect on System Performance}

	\descri 
		Running the logging code and writing log entries introduces \textit{performance overhead}. This may be expected, or it may be \textit{parasitic} when logging is disabled, but the logging code is still executed without producing any output.
	
	\causes 	
		The performance overhead becomes problematic when the volume of logging is high, as in ``Landfill Logs'' (LS7). In such cases, the cost is visible in the logs, while parasitic overhead caused by poorly configured guards, as in ``Sleeping Guards'' (LS8), is more difficult to detect. Reduced code efficiency associated with ``Logging Code Debt'' (LS9) can also contribute to increased overhead.
	
	\implications
		The performance impact of logging can appear in many parts of the system, including CPU time, memory usage, disk/IO bandwidth, storage, battery consumption, and response time, ultimately affecting the end‑user experience. Developers may not be fully aware of these costs, and determining the optimal amount of logging is difficult because system workloads vary.
	
	\references{ 
		\cite{ Bogatinovski2023Auto}, \cite{ chen_characterizing_2017}, \cite{ chen_extracting_2019}, \cite{chen_survey_2022}, \cite{ding_log2_nodate}, \cite{fu_where_2014}, \cite{gu_logging_2023}, \cite{ Kulkarni2024}, \cite{ li_qualitative_2021}, \cite{ li_guiding_2020}, \cite{ oliveira_automatically_2020}, \cite{ Lu2018CloudRaid}, \cite{Mace2018Pivot}, \cite{ marron_log_2018}, \cite{ rong_how_2023}, \cite{ Wang2025}, \cite{ Yu2023LogReducer}, \cite{ zeng_studying_2019}, \cite{zhao_game_2017}}

\smelltitle{CO4}{Accidental Side-effects of Logging Code}

	\descri
		 Executing logging code can change the state of the software.
	
	\causes
		This issue may be caused by ``Landfill Logs'' (LS7), as unnecessary logging code can reduce code comprehensibility. Likewise, the ``Logging Code Debt'' (LS9) smell negatively affects code clarity, making accidental side‑effects—such as unsafe variable usage—more likely.
	
	\implications	
		Logging code may unintentionally change program state, resulting in side-effects that cause bugs or crashes. These issues are often hard to locate and can impact both developers and end‑users.
	
	\references{ 
		\cite{ batoun_literature_2024}, \cite{ chen_extracting_2019}, \cite{ Kulkarni2024}, \cite{ marron_log_2018}, \cite{ Wang2025}, \cite{ yuan_zing_2012}}
		
	\example 
		Unexpected behavior may occur in a multithreaded program, as deadlocks can arise if the logging mechanism is not thread‑safe. A real‑life example of a misplaced logging statement that corrupts program state is shown in~\cite[Fig.~3]{yuan_zing_2012}

\setlength{\parindent}{1pc}  \setlength{\parskip}{0pt}

\subsubsection{Taxonomy Validation}
\label{sec:taxonomy_validation}

The independent classifications (see Step~13 ``validation'' in~\S~\ref{sec:taxonomy_creation}) produced a Cohen's kappa of $\kappa = 0.83$.
Table~\ref{tab:taxonomy_validation} shows how the issues in the validation dataset were classified according to the proposed taxonomy. Because a single issue may exhibit multiple log smells, the counts are not mutually exclusive.

\begin{table}[t]
	\centering
	\caption{Distribution of log smells identified in the Defects4Log validation dataset. A single issue may be associated with multiple log smells.}
	\label{tab:taxonomy_validation}
	\begin{tabular}{llrr}
		\hline
		\textbf{ID} &
		\textbf{Log smell} &
		\textbf{\# Issues} &
		\textbf{Percentage (\%)} \\
		\hline
		LS1 & Format Drift             & 16 & 8.9\% \\
		LS2 & Undercover Identifier    & 0 & 0.0\% \\
		LS3 & Shifting Logging Levels  & 31 & 17.3\% \\
		LS4 & Deceptive Variable       & 57 & 31.8\% \\
		LS5 & Misleading Messages      & 40 & 22.3\% \\
		LS6 & Logging Blackout         & 0 & 0.0\% \\
		LS7 & Landfill Logs            & 12 & 6.7\% \\
		LS8 & Sleeping Guards          & 0 & 0.0\%\\
		LS9 & Logging Code Debt        & 20 & 11.2\% \\
		-- & Something Else        & 3 & 1.7\% \\
		\hline
		\multicolumn{2}{l}{\textbf{Total}} & 179 & \\
		\hline
	\end{tabular}
\end{table}

The distribution shows that only three of the 164 logging issues could not be mapped to the proposed taxonomy. The remaining 161 (98.2\%) issues were assigned to one or more log smells, suggesting that the taxonomy captures a broad range of logging issues observed in practice. 
``Misleading Messages'' (LS5), ``Deceptive Variable'' (LS4), and ``Shifting Logging Levels'' (LS3) accounted for 72\% of the issues. 
Their higher frequency may be partly explained by the fact
that these smells can often be identified directly from
logging statements in the source code.

Meanwhile, three smells were not observed in the dataset. ``Undercover Identifier'' (LS2) concerns the 
information available in the resulting log entries and therefore cannot generally be identified from 
logging code changes alone. Regarding ``Logging Blackout'' (LS6), the dataset did not contain 
cases in which developers introduced entirely new logging statements. Finally, for ``Sleeping Guards'' (LS8), 
although the dataset contained modifications to existing logging guards, such as the example shown in Figure~\ref{fig:example_code_change}, 
we did not observe cases in which the primary issue was the absence or misuse of a logging guard.

\begin{figure}[t]
	\centering
	\begin{lstlisting}[
		language=diff,
		numbers=left,
		numberstyle=\small\color{gray},
		numbersep=10pt,
		xleftmargin=25pt,
		frame=none
		]
if (saslServer.isComplete()) {
-     LOG.info('SASL server context established. Negotiated QoP is '
-		+ saslServer.getNegotiatedProperty(Sasl.QOP));
+     if (LOG.isDebugEnabled()) {
+         LOG.debug('SASL server context established. Negotiated QoP is '
+			+ saslServer.getNegotiatedProperty(Sasl.QOP)); }
		\end{lstlisting}
		\caption{Example of a code change reducing excessive logging output by changing the logging level and introducing a logging guard.~\protect\footnotemark.}
		\label{fig:example_code_change}
	\end{figure}
	\footnotetext{\url{https://issues.apache.org/jira/browse/HADOOP-7272}}

The three issues that could not be classified using the taxonomy addressed concerns outside the scope of the identified log smells. One concerned transaction logs rather than application logging. Another involved concurrency and the execution context of the logging code rather than the quality of the logging itself. The third addressed the use of string concatenation instead of the logging framework's parameterized message mechanism, resulting in a performance concern rather than a log smell.

During the discussions to reach consensus on the classifications, most disagreements arose when a change could reasonably be associated with multiple log smells. For example, changes reducing excessive logging output could be interpreted either as ``Shifting Logging Levels'' (LS3) or as ``Landfill Logs'' (LS7), depending on whether the emphasis was placed on the logging level or the amount of generated log data. An example of such a change is shown in Figure~\ref{fig:example_code_change}, where the description was: \say{Two info logs are printed when connection to RPC server is established, is not necessary. On a production cluster, these log lines made up of close to 50\% of lines in the namenode log.}. Similar disagreements concerned the distinction between ``Misleading Messages'' (LS5) and ``Format Drift'' (LS1). Furthermore, some code changes simultaneously affected multiple logging elements, such as logged variables, static messages, and logging levels. In these cases, the disagreement was primarily about which affected logging element represented the primary manifestation of the issue a

At the facet level, disagreements also concerned
distinctions among incorrect, imprecise, and
linguistically problematic log messages. These
observations suggest that assigning facets may
require additional contextual information and
interpretation of developers' intent. Therefore,
we focus the inter-rater agreement analysis
on the log smell level.

Overall, the validation results suggest that the proposed taxonomy provides broad coverage of logging issues while highlighting the importance of identifying the primary manifestation of issues that affect multiple logging elements.

\subsection{RQ2: Tools repairing log smells}
\label{sec:RQ2_results}

In RQ1, we defined several log smells, and in this RQ, we focus on identifying existing repair tools for them. Although it is unlikely that all smells can be fully addressed automatically, many of them could be mitigated using repair tools. We define a \textit{repair tool} as a tool or method that detects an issue and proposes a modification intended to remove or repair it.
From the papers included in our survey, we identified 16 repair tools
targeting log smells and five tools addressing other logging
issues. The repair tools are described in
Table~\ref{tab:repair_tools}. Because our focus is on log smells, the
other tools are described in the replication package.

\begin{table}
	\caption{Tools repairing log smells.}
	\begin{tabular}[h!]{p{2cm}p{12cm}l}
		\toprule
		\textbf{Tool} & \textbf{Description} & \textbf{Ref} \\
		\midrule
		DeepLV & Provides recommendations for appropriate logging levels; it is a deep-learning-based method using the syntactic context and message features. &  \citep{li_deeplv_2021} \\
		Errlog &  Proactively inserts logging statements into the source code, particularly for unlogged exceptions. It uses the Saturn framework~\citep{aiken2007saturn} to implement a novel algorithm that identifies potential exception locations and checks if they are already handled, while minimizing performance overhead. & \citep{yuan_be_nodate} \\
		He2018 & Adds logging descriptions based on existing ones; it is based on information retrieval. & \citep{he_characterizing_2018-1} \\
		Kim2020 &  Assesses the appropriateness of log levels and recommends alternative levels if necessary. It uses machine learning methods that learn the semantic and syntactic features of log messages. & \citep{kim_automatic_2020} \\
		LACC &  Predicts the location and severity of missing log statements in code clones. It detects methods with logging code, calculates log-aware source code features for them, and finally makes predictions using machine learning. & \citep{gholamian_logging_2020} \\
		Li2017 &  Suggests a log level for new logging statements. It uses ordinal regression models to learn the levels from the project history. & \citep{li_which_2017} \\
		Li2021 & Suggests logging for exception stack traces. Predictions are made using machine learning models that are trained with several code metrics. & \citep{li_studying_2021}  \\
		Liu2020 & Automatically generates descriptive texts for logging statements, reducing the problem to a retrieval-based Q\&A task. &  \citep{oliveira_automatically_2020} \\
		Liu2021 & Recommends which variables to log and is also able to consider out-of-vocabulary words. The tool addresses this as a representation problem and trains a neural network for presenting program tokens and based on that predicts whether an identifier should be logged. &  \citep{liu_which_2019} \\
		LogReducer & detects the most recurring log templates and reduces them on the fly. It is a non-intrusive and language independent framework based on extended Berkeley Packet Filter. &  \citep{Yu2023LogReducer}  \\
		LogUpdater &  Detects and updates logging statements for statement code inconsistency, static dynamic inconsistency, temporal relation inconsistency and readability issues. It is a two-phase framework that utilizes an LLM. & \citep{Zhong2025LogUpdater} \\
		LTID & Propagates missing IDs to logs. The method uses log graphs and ID variables in log templates. &  \citep{zhao_studying_2023}  \\
		PADLA & Dynamically adjusts the logging verbosity level of a running system to better record irregular events. It is an extension of the Log4j logging framework employing an online phase detection algorithm. &  \citep{mizouchi_padla_2019} \\
		QuLog & Evaluates the correct log level and detects if the messages have a sufficient linguistic structure. It utilizes deep learning and explainable AI methods. &  \citep{bogatinovski_qulog_2022}  \\
		TeLL & A log level predictor that considers both intra-block and inter-block information for the prediction. It utilizes graph neural networks. &  \citep{Liu2022}  \\
		Verbosity\-Level\-Director & Provides guidance on certain types of problematic verbosity levels. The tool builds a prediction model using features related to triggered methods, logging content, exception type, post‑processing, and code comments. &  \citep{anu_approach_2019} \\
		\bottomrule
	\end{tabular}
	\label{tab:repair_tools}
\end{table}

To understand which log smells the identified tools address,
Table~\ref{tab:tools_repair_results} maps the tools to the smell
facets they repair. If a tool repairs a facet, it is listed in the
\emph{Tool} column; facets without an identified repair tool are
marked with ``-''. Some tools address only a subset of a facet; 
these sub-facets are indicated using an asterisk (*).

The first observation is that six out of the nine smells are covered
by at least one repair tool. However, the coverage is uneven as the
number of tools for a smell varies between zero and
seven. Furthermore, the identified tools only address 15 out of the 29
identified facets. This suggests that \textit{log smells are
  moderately covered by repair tools, but further research is needed.}

The tools appear to focus primarily on issues visible in log entries,
such as logging levels, messages, and logged variables. In contrast,
code-related smells receive considerably less attention. One possible
explanation is that tools addressing issues related to logging code
may be developed for other software engineering tasks and are
therefore not explicitly presented as logging tools. For example,
several tools have been developed for detecting and fixing code
smells~\citep{lacerda_code_2020}. As a result, the ``Logging Code
Debt'' (LS9) log smell, i.e., issues related to the \textit{code} used
to emit logs, is likely already at least partially addressed by
existing code maintenance tools, which were not within the scope of
this study.

The focus on the log-entry level is also reflected in the most
frequently addressed smell, as seven tools target ``Shifting Logging
Levels'' (LS3). Furthermore, six of these tools focus on the
``incorrect (logging level)'' facet. This suggests that assigning
appropriate logging levels has received considerable attention from
the research community, while the ``inconsistent (logging level)''
facet has been studied considerably less. One possible explanation is
that logging levels are often assigned in an ad-hoc manner and are
frequently revised during software
evolution~\citep{yuan_zing_2012}. As a result, incorrect logging
levels might be perceived as a common practical problem. Moreover, the
identified tools indicate that this facet is well suited for automated
analysis, as several approaches use statistical and machine learning
models to predict suitable logging levels.

Most of the addressed smells are targeted by multiple tools. However,
this does not necessarily imply competition between the tools, as many
approaches are complementary rather than redundant. For example, for
``Logging Blackout'' (LS6) and ``Landfill Logs'' (LS7), the tools
target different facets.
Despite this, we identified three facets that are addressed by
multiple tools: six tools address ``incorrect (logging level)'', while
two tools address ``missing variable'' and ``missing/inconsistent
message''. Even in these cases, substantial differences exist between
the approaches. For example, tools targeting incorrect logging levels
rely on different sources of information, including code and project
metrics, code clones, and logging context. Consequently, the tools may
identify different entries as incorrect or suggest different
repairs. Furthermore, some tools address only a subset of a facet. For
instance, Errlog focuses on missing logging statements in exception
code, whereas other approaches aim to provide more general logging
recommendations. Therefore, overlap in the targeted facet does not
necessarily imply overlap in functionality, and multiple tools may be
used in a complementary manner.

While some smells are addressed by several tools, we did not identify
tools for some smells. First, ``Format Drift'' (LS1) was not addressed
by any of the identified tools. This was somewhat unexpected, given
the importance of consistent log formats for automated log analysis
and the extensive research on log parsing. One possible explanation is
that the research has primarily focused on developing parsing
techniques that can tolerate diverse log formats, rather than tools
that repair format inconsistencies~\cite{zhu_tools_2019}. In addition, libraries often automatically
enforce log formatting, which may reduce the practical need for
dedicated repair tools. Finally, we excluded solutions that propose
new logging approaches rather than repairing specific logging
issues. Such systems, including Log++~\citep{marron_log_2018}, often
promote uniform logging practices and may therefore indirectly
mitigate LS1, even though we did not classify them as repair tools.

Similarly, no repair tools were identified for ``Sleeping Guards''
(LS8), despite several works proposing techniques for detecting the
smell~\citep{hassani_studying_2018, zhi_towards_2022}. One plausible
explanation is that automatically repairing logging guards is more
difficult than detecting them. Addressing missing or incorrect
guards may require reasoning about control flow and the intended
purpose of the logging statement. This gap highlights an opportunity
for future work to develop automated repair techniques for LS8.

More broadly, many facets lacking repair tools concern what information should be logged and how it should be represented. Examples include incorrect identifiers, incorrect variables, imprecise messages, and incorrect logging guards. Unlike facets such as incorrect logging levels, for which patterns can be learned from source code and logging context, these facets often require a deeper understanding of the application's behavior and the intended purpose of the logging statement. This may be one reason why these kinds of repair tools appear to be missing.

\begin{table}
	\centering
	\small
	\caption{Mapping between facets of log smells and tools repairing
		them.}
	\label{tab:tools_repair_results}
	\begin{tabular}{lL{3.1cm}L{6.1cm}L{4.4cm}}
		\toprule
		\textbf{ID}	&	\textbf{Log smell}	&	\textbf{Facets}	&	\textbf{Tool}	\\	\midrule
		LS1	&	Format Drift	&	Incomplete, ambiguous, not following a format, multiple formats	&	 -	\\
		\rowcolor{gr}
		LS2	&	Undercover identifier	&	Missing	&	LTID	\\ \rowcolor{gr}
				&		&	Wrong	&	 -	\\ 
		LS3	&	Shifting Logging Levels	&	Incorrect	&	DeepLV, Kim2020, Li2017, LACC, QuLog, TeLL	\\
				&		&	Inconsistent 	&	VerbosityLevelDirector	\\
				&		&	Missing	&	 -	\\
		\rowcolor{gr}
		LS4	&	Deceptive Variable	&	Missing	&	Liu2021, LogUpdater	\\ \rowcolor{gr}
				&		&	Wrong	&	 LogUpdater	\\ \rowcolor{gr}
				&		&	Malformed output, varying granularity	&	 -	\\
		LS5	&	Misleading Messages	&	Missing, inconsistent (adds descriptions)	&	He2018, Liu2020	\\
				&		&	Wrong, language issues	&	 LogUpdater	\\
				&		&	Imprecise, duplicated	&	 -	\\
				\rowcolor{gr}
		LS6	&	Logging Blackout	&	Overlooked &		\\ \rowcolor{gr}
				&		&	* Exceptions	&	Errlog	\\ \rowcolor{gr}
				&		&	Hidden	&	PADLA	\\
		LS7	&	Landfill Logs	&	Too detailed	&	PADLA	\\
				&		&	Redundant, Useless	&	LogReducer	\\
				&		&	Stack trace overuse	&	Li2021	\\
		\rowcolor{gr}
		LS8	&	Sleeping Guards	&	Missing, incorrect	&	-	\\
		LS9	&	Logging Code Debt 	&	Comprehensibility, long and duplicated code	&	-	\\
		\bottomrule
	\end{tabular}
\end{table}

 \section{Discussion}
\label{sec:discussion}

The concept of ``smell'' is widely adopted in software engineering to describe potentially harmful practices. Understanding such smells helps developers improve software quality, motivating researchers to create taxonomies for different smell types. Although several studies discuss logging issues, we identified only two previous works that have explicitly used the term ``log smell''. Despite providing a general definition of log smells, both of these studies focused on specific smells rather than investigating log smells in general or providing a comprehensive taxonomy.

In this paper, we have conducted a literature survey and---based on
the logging issues identified through the survey---created a taxonomy of nine log smells and identified five causes and four consequences associated with
them. Each smell corresponds to a different aspect of logging and can be manifested by multiple facets. To assess whether the taxonomy is applicable beyond the surveyed literature, we performed a preliminary validation using 164 real-world logging issues extracted from open-source projects. The validation showed that the vast majority of the issues could be mapped to the proposed taxonomy.
The need to address logging issues has long been recognized in the research community, and
the topic has been extensively researched. Therefore, we have examined
whether the logging tools introduced by the research community address
the identified smells and identified 16 tools designed to repair them.

The survey investigated logging-related issues reported in the surveyed literature, revealing \emph{a wide variety of log smells
that manifest in different aspects of logging}. Regardless of the smell, if emerging issues are not taken seriously, they might create or evolve into other problems. Therefore, developers should be aware of these smells and their effects on logging quality. 
However, some of the smells can have a more widespread negative impact than others. For example, ``Landfill Logs'' (LS7) can affect system performance, storage costs, and log comprehensibility; this smell was considered a potential cause of all four identified consequences of log smells. On the other hand, ``Misleading Messages'' (LS5) mainly affects the usefulness and comprehensibility of the log entries and was identified as a potential cause only once. Three smells (LS1, LS4, LS6) were not determined to be the cause of any of the consequences. However, this does not imply that these are not potentially harmful. The paper focused on direct consequences, and while none of these smells might directly cause a significant problem, they all hinder the comprehensibility, completeness, and thus usefulness of the log, making the downstream tasks more difficult.

\label{discussion:rq2}
Our analysis of repair tools reveals a clear imbalance in the
  current state of automated support for log smells. While six of the
  nine identified smells are addressed by at least one repair tool,
  tool coverage remains limited at the facet level, with only 15 of
  the 29 identified facets supported. Moreover, the available
  approaches concentrate primarily on issues visible in individual log
  entries, such as logging levels, messages, and logged variables. In
  contrast, smells related to log structure (LS1) and logging code
  (LS8 and LS9) have received comparatively little attention. Furthermore, the facets that were unaddressed by existing tools suggest that current research is primarily focused on aspects of logging that can be inferred directly from source code, log messages, or project history, whereas issues requiring a deeper understanding of the software behavior, execution context, or developer intent seem more difficult to repair automatically.

At the same time, smells not addressed by repair tools should not automatically be interpreted as unsupported in practice. Some smells, particularly ``Logging Code Debt'' (LS9), overlap with software maintenance. Existing code smell detection, refactoring, and program comprehension tools may therefore already mitigate parts of these smells, even if they were not designed specifically for logging. This highlights an opportunity for future work not only to develop new logging-specific repair techniques, but also to evaluate the extent to which existing software quality tools improve logging code.

Our findings are in line with recent studies on logging. In their
systematic mapping study, \citet{gu_logging_2023} concluded that one
of the aspects requiring more research focus is the degree to which the intentions and concerns of
logging can be implemented. They also noticed that most tools are
focused on the ``where and what to log'' categories. Our results are 
similar, as we identified several tools addressing issues related to where to log;; the identified log smells indicate that many aspects of ``how
good logging is'' still require more attention. In an interview at Microsoft, \citet{he_empirical_2022} reported that developers expressed a need for off-the-shelf logging tools to assist them. This is also visible in our results, as even though several tools exist, many address only one logging issue. This makes adopting them more difficult, as developers would need to use several tools to address all relevant logging issues.

We acknowledge that we have defined smells at a higher level than in some other areas of software engineering. For instance, in our taxonomy, duplicated logging code is categorized under the log smell ``Logging Code Debt'', which contains general logging code-related issues. This smell includes code smells, with ``Duplicated Code'' being a code smell on its own. However, we found that using higher-level groupings and then providing the facets for each smell is more practical and easier to understand than listing 29 separate smells.

 \section{Threats to Validity}
\label{sec:threats_to_validity}

\textbf{Internal validity} is related to the factors that affect the data and the inferences made about them. 
Because we did not conduct an SLR and did not apply snowballing, some logging issues reported in the literature may have been missed, potentially affecting the completeness of the resulting taxonomy. To mitigate this threat, we used multiple search queries across several data sources, followed SLR guidelines where applicable, and provide a comprehensive replication package. Furthermore, the independent validation described in Section~\ref{sec:data_analysis} identified additional issue descriptions but did not require modifications to the taxonomy structure or introduce new log smells, causes, or consequences. Therefore, while we do not claim that the taxonomy is exhaustive, these observations increase our confidence that it provides a robust synthesis of the logging issues identified in the surveyed literature.

Results related to tools may also be incomplete. Our queries did not include tool‑specific terms, which may have caused us to miss relevant papers. We mitigated this by including tools reported in the reviewed secondary studies, several of which our search had not identified. In addition, some facets might be addressed by tools from other software engineering domains (e.g. code smell tools). Because our survey focused on logging, we did not include such tools.

Data extraction and taxonomy construction are subjective and may therefore introduce errors such as missed or misinterpreted information. To mitigate this threat, Authors 1 and 2 independently screened the query results and Cohen's Kappa for paper inclusion between them was 0.807. While the taxonomy was based on the data extraction of Author 1, Author 2 independently extracted quotations from the included papers, derived the corresponding logging issues without consulting Author 1's issue list, and assessed whether the additional issues could be mapped to the taxonomy. While this validation resulted in several refinements to issue descriptions, facets, and examples, it did not require structural changes to the taxonomy.

Classifying log issues relies on domain expertise and, consequently, different researchers might categorize issues differently, affecting both the taxonomy and the identified tools. For example, tools addressing ``where to log'' were excluded even though they could be interpreted as addressing ``Logging Blackout'' (LS6), and tools proposing entire logging systems were also excluded although they might address multiple smells. We mitigated this by conducting a card sort involving multiple authors, recording agreement before discussion, and iteratively refining the classification until consensus was reached.

Publication bias is a common threat in literature surveys. We consider it minor here because our selection criteria did not require papers to propose solutions. Instead, the criteria aimed to capture a representative set of logging issues, without strict exclusions based on venue or paper type.

\textbf{Construct validity} is related to the degree to which the adopted measures represent the theoretical constructs investigated in the study. Because the notion of a smell is well‑established, we defined log smells based on existing smell definitions in software engineering and previous logging research.

\textbf{External validity} relates to the generalizability of the results. Our goal was to collect a set of logging issues; therefore, variation in study quality was not considered as a major threat. The exclusion of gray literature may limit coverage, as some issues may be known only in industry~\citep{GAROUSI2019101}. However, the resulting taxonomy synthesizes a broad range of issues reported in the surveyed literature and can be extended or refined as new issues emerge.

The taxonomy validation focused on a single dataset: the Defects4Log dataset~\cite{Wang2025}. Since the corresponding study was included in our literature survey, the validation is not fully independent from the taxonomy creation. However, the dataset was derived independently from issue reports and commit histories of multiple open-source projects. Therefore, we consider it suitable for an initial assessment of whether the identified log smells can be observed in practice. 
As the dataset primarily contains logging code
changes, it provides stronger evidence for log smells that
can be identified from source code than for those that
require examining the generated log artifacts. 
Furthermore, the validation was performed by Authors~1 and~2, who contributed to the development of the taxonomy and may therefore have been influenced by prior knowledge of it. To mitigate this threat, the issues were classified independently by both authors, and the observed agreement was high ($\kappa = 0.83$).

\textbf{Reliability} concerns the trustworthiness and repeatability of the results. To support replication, we documented all methodological steps and included the full results from all steps in the replication package.

 \section{Related Work}
\label{sec:related_work}
Several previous works have described and categorized logging
issues. To better position our contribution, we have mapped the issue
categories presented in prior studies to the log smells in our
taxonomy; this mapping is presented in
Table~\ref{tab:taxonomy_mapping}.
Column ``Mapping Type'' describes the relationship between the issue
presented in the compared study (``Presented Issue'') and the mapped
element in our taxonomy (``Mapped Element''). Four mapping types are
used:
\begin{itemize}
\item ``Same'' indicates that the presented issue corresponds directly
  to a log smell in our taxonomy;
\item    ``Facet'' indicates that the presented issue corresponds to a
  single facet of a smell or a category;
\item  ``Subset'' indicates that the presented issue covers multiple
  facets of a smell or a category but does not cover it completely;
\item 
 ``Generalization'' indicates that the presented issue is a broader
 category that overlaps with one or more smells or categories but does
 not distinguish between their individual manifestations.
\end{itemize}

The table does not include all papers that mention logging issues, as this would encompass all of the papers in our survey. Instead, we focus on studies that propose categorizations of logging issues or investigate complete log smells. Therefore, we exclude studies that address only individual facets of a log smell as well as secondary studies whose primary goal is to survey the logging literature rather than categorize logging issues, such as the study by~\citet{batoun_literature_2024}, which synthesizes logging research and practitioner concerns but does not classify logging issues in a way that is comparable to ours.

\begin{table}
	\caption{Mapping issue categories from prior logging studies to the proposed taxonomy. The mapping type describes how each issue category relates to the mapped element.}
	\label{tab:taxonomy_mapping}
	\footnotesize
	\centering
\begin{tabular}{p{0.8cm}p{6.6cm}p{2.9cm}p{2cm}}
	\toprule
	\textbf{Ref.} & \textbf{Presented Issue }&  \textbf{Mapped Element} & \textbf{Mapping Type} \\
	\midrule
	
	\multirow{6}{*}{\cite{chen_characterizing_2017}}
	& Nullable Objects & CO4 & Facet \\
	& Explicit Cast & CO4 & Facet \\
	& Wrong Verbosity Level & LS3 & Facet \\
	& Duplication with the Definition of Another Method & LS9 & Facet \\
	& Duplication with the Definition of a Local Variable & LS9 & Facet \\
	& Malformed Output & LS4 & Facet \\
	
	\midrule
	
	\multirow{8}{*}{\cite{gu_logging_2023}}
	& Lacking crucial messages & LS6 & Generalization \\
	& Redundant or useless messages & LS7 & Same \\
	& Incorrect or ambiguous messages & LS5 & Generalization \\
	& Leakage of sensitive data & CO1 & Same \\
	& Maintenance barriers & CA5 & Generalization \\
	& Log verification and validation difficulties & Other (Validating and verifying logs) & Same \\
	& Performance overhead & CO3 & Same \\
	& Heterogeneity of the log messages & LS1 & Generalization \\
	
	\midrule
	
	\multirow{7}{*}{\cite{hassani_studying_2018}}
	& Inappropriate log messages & LS5 & Generalization \\
	& Missing logging statements & LS6 & Generalization \\
	& Inappropriate log level & LS3 & Generalization \\
	& Log library configuration issues & CA3 & Generalization \\
	& Runtime issues & CO4 & Generalization \\
	& Overwhelming logs & LS7 & Same \\
	& Log library changes & CA3 & Facet \\
	
	\midrule
	
	\multirow{4}{*}{\cite{he_empirical_2022}}
	& Too many logs & LS7 & Same \\
	& Missing logs & LS6 & Same \\
	& Hard to find failure-related logs & LS7 & Facet \\
	& Logs are noisy & LS7 & Facet \\
	
	\midrule
	
	\multirow{1}{*}{\cite{li_studying_2022}}
	& Duplicate Logging Statements & LS3, LS5, LS6, LS9 & Generalization \\
	
	\midrule
	
	\multirow{13}{*}{\cite{patel_sense_2022}}
	& Missing variables and statements & LS4, LS5, LS6 & Generalization \\
	& Imprecise logging messages & LS5 & Subset \\
	& Null pointer dereference & CO4 & Facet \\
	& Uninitialized variables & CO4 & Facet \\
	& Logging wrong variables & LS4 & Facet \\
	& Fixing incorrect log levels & LS3 & Facet \\
	& Fixing inconsistencies & LS5, LS9, CA3 & Generalization \\
	& Deleting redundant information & LS7 & Facet \\
	& Language mistakes & LS5 & Facet \\
	& Fixing format specifiers & LS4, CO4 & Subset \\
	& Message mistakes & LS5 & Facet \\
	& Formatting issues & LS5 & Facet \\
	& Revealing kernel pointers & CO1 & Facet \\

	\midrule
	
	\multirow{7}{*}{\cite{Wang2025}}
	& Readability Issues & LS5 & Subset \\
	& Variable Issues & LS4, CO4 & Subset \\
	& Logging Level Issues & LS3 & Facet \\
	& Semantics Inconsistent with Context & LS4, LS5 & Subset \\
	& Sensitive Information & CO1 & Subset \\
	& Insufficient Information & LS4, LS5, LS6 & Generalization \\
	& Performance Issues & CO3 & Subset \\
	
	\midrule
	
	\multirow{3}{*}{\cite{yuan_zing_2012}}
	& Verbosity-level modifications & LS3 & Facet \\
	& Variable-logging modifications & LS4 & Subset \\
	& Static-content modifications & LS5 & Generalization \\
	
	\midrule
	
	\multirow{1}{*}{\cite{zhao_studying_2023}}
	& Missing identifiers in logging statements & LS2 & Generalization \\
	
	\midrule
	
	\multirow{1}{*}{\cite{zhi_towards_2022}}
	& Missing Logging Guards & LS8 & Same \\

	\midrule
	
	\multirow{4}{*}{\cite{Zhong2025LogUpdater}}
	& Statement-code inconsistency & LS5 & Facet \\
	& Static-dynamic inconsistency & LS4 & Subset \\
	& Temporal inconsistency & LS5 & Facet \\
	& Readability issues & LS5 & Facet \\
	
	\bottomrule
\end{tabular}
\end{table} 

Table~\ref{tab:taxonomy_mapping} shows that the issues represented by the log smells in our taxonomy have all been discussed previously in the literature. However, this knowledge is scattered across multiple studies and often appears under different names, at different levels of granularity, or as part of broader categorizations of logging issues. The table also shows that prior categorizations cover different subsets and perspectives of logging issues, whereas our taxonomy combines these issues into one work. While some of the proposed smells correspond closely to issue categories identified in earlier work (e.g., LS8 ``Sleeping Guards''), others consolidate related issues that have previously been discussed independently (e.g., LS5 ``Misleading Messages'').

While Table~\ref{tab:taxonomy_mapping} compares our taxonomy against prior categorizations of logging issues, the remainder of this section discusses broader secondary studies on software logging and prior work specifically addressing the notion of log smells.

\subsection{Secondary studies on software logging}
\label{sec:RW_SLRs}

Logging issues have been investigated by several recent works, with the most similar to our work being the systematic mapping study by~\citet{gu_logging_2023}. They investigated 56 papers to identify major issues in logging, their solutions, and gaps in research. They identified eight main issues and categorized them into \textit{where to log, what to log, why to log}, and \textit{how well is the logging}. 
The study found no issues related to the why-to-log category, and most  solutions addressed where-to-log or what-to-log issues. This highlights gaps in understanding why and how well logging is performed. Although the authors noted that research on logging remains active, many issues are still unresolved. The key difference from our work is in the perspective: while their study examines logging through the lens of where, what, why, and how to log---emphasizing the logging process---our work focuses specifically on logging issues, and in particular on log smells.

\citet{MADI2026107961} conducted a systematic review of 21 studies on logging‑smell detection, classifying the tools using the taxonomy introduced in the pre‑print version of this article~\citep{saarimaki2024taxonomysoftwarelogsmells}. They found that although every smell is covered by at least one detection tool, most focus on logging levels and log messages. 
The review also indicated that research remains fragmented, with studies targeting different smells, using inconsistent terminology, and relying on non‑standardized datasets and evaluation methods. The authors identify recurring challenges---including the lack of logging guidelines, data imbalance, and limited cross‑project generalizability. They also observe a shift from static analysis and traditional ML toward deep learning and LLM‑based techniques in more recent work.

A systematic literature review by \citet{batoun_literature_2024} inspected the state-of-the-art logging practices, focusing on instrumentation, storage, and preprocessing. The study examined both the academic literature and a practitioner Q\&A forum (StackOverflow). The authors identified a gap between research and practice: five of the seven high‑level topics raised by practitioners—including context‑dependent log usage—were absent from the scientific literature. 

A survey of 112 papers by \citet{gholamian_comprehensive_2022} examined the state of the art in logging research, identifying the main areas studied and proposing a taxonomy of the field. Similarly to our work, the authors assess the costs, i.e., issues, of logging. Although the work includes several issues such as performance cost and noisy log files, this was not the main aspect of the study. Finally, the work presents several aspects of logging that require more research, such as cost-aware logging, log maintenance, and improved logging practices.

\citet{zhaoxue_survey_2021} surveyed log enhancement, log parsing, and log analysis in the context of AIOps and big data. Our work aligns with their focus on log enhancement, particularly in determining what and where to log. They highlight challenges similar to ours and additionally identify several trends in log‑enhancement research, including the use of increasingly fine‑grained features, coupling of subsequent processes, and automation to avoid inconsistent logging.

\citet{chen_survey_2022} conducted a survey of 69 papers on log
instrumentation, i.e., inserting logging code. The work focuses on
logging approaches, logging utility integration, and logging code
composition. The work identified nine challenges, which were classified
into usability, diagnosability, logging code quality, and security
compliance-related issues. In addition to the issues, the work also
discusses the solutions.

\citet{he2021survey} provide a comprehensive overview of automated log‑analysis research, covering logging, log compression, parsing, mining, open-source toolkits and datasets, and best practices. Their goal is to help practitioners understand available techniques along with their benefits and challenges. Although they discuss logging challenges---organized around where, what, and how to log---and mention work on detecting anti‑patterns, the paper primarily focuses on log analysis rather than logging issues.

\citet{candido_log-based_2021} conducted a systematic mapping study examining challenges across the log data life-cycle. They organized the findings into three areas: logging, log infrastructure, and log analysis. From the 24 logging‑related papers, they concluded that developers need better logging tools, but unclear logging requirements hinder tool development.

\citet{Wang2025} investigate logging code defects through a systematic
literature review combined with the analysis of issue reports and
logging-related commits. They propose a taxonomy of seven logging code
defect patterns: readability issues, variable issues, logging level
issues, semantic inconsistencies between logging statements and code
context, sensitive information leakage, insufficient information, and
performance issues. In addition, they define 14 scenarios that
describe how these defect patterns manifest in practice.
The paper is included in our survey, and therefore its defect patterns are covered by our taxonomy. However, they describe these issues at a finer granularity by defining specific defect patterns and scenarios, whereas our facets capture broader manifestations of the same problems. Therefore, their work is complementary to ours.

\subsection{Prior Work on Log Smells}
\label{sec:rw_log_smells}

To the best of our knowledge, log smells have been defined in two previous works.

\citet{li_dlfinder_2019} investigate duplicate logging code smells, i.e., logging statements that have identical static messages. In their study, they identify five harmful patterns of duplicate messages as well as six justifiable patterns that are not considered problematic. Following prior code-smell studies, the authors define log smells as \say{surface indication that usually corresponds to a deeper problem in the system}. Despite the general nature of this definition, the study does not address log smells broadly, but instead provides an in-depth analysis of a single, specific log smell related to duplicate logging statements in source code.

The other work adopting the term log smell is by \citet{chen_characterizing_2017}. The paper investigates anti-patterns in logging code and defines six logging anti-patterns, including a category ``Logging Code Smells''. Following the general notion of software smells, the authors define log smells as \say{poor design and implementation choices when developing logging code}. However, they provide only a single concrete example of such issues, namely the long logging snippet.

Both studies ground their definition of log smells in established software smells. However, both studies focus on a specific instance of log smell rather than providing a broader characterization. In particular, both works restrict log smells to issues in logging code and do not consider problems manifesting in the resulting log artifacts. Our work builds on these definitions by broadening the notion of log smells and explicitly accounting for the dual nature of logging, covering issues that affect both logging code and log files.

\subsection{Issues in non-software-based logs}

Several log types exist beyond software logs, including event and security logs~\cite{alma9922858298007251}. Event logs are the closest to software logs and are used in process mining; they record executed tasks, timestamps, and more. Because they are created for a specific downstream task---often from multiple sources---they contain more refined information than raw software logs.

Several studies have examined issues in event logs and proposed tools to address them. \citet{Comuzzi2025} identify 11 imperfection patterns in event logs and introduce a formal language to describe problems in event log data quality. Many of these issues also apply to software logs; however, there are differences stemming from the differences in the nature of the logs. For example, logs related to a specific task being scattered is an issue as it hinders the analysis task, but it is not an issue per se for software logs. 

Similarly,~\cite{SURIADI2017132} present 11 event log imperfection patterns. They describe the patterns as well as provide additional information such as a strategy to detect and rectify them. The patterns are based on the experiences of the authors and were validated using real event logs. 
\cite{Fischer2020} conducted a study solely focusing on timestamp-related issues in event logs. They define 15 quality metrics and propose an approach to detect such issues.

 \section{Conclusion}
\label{sec:conclusion}

In this work, we have conducted a literature survey and used the identified logging issues to derive a taxonomy of \emph{log smells}, comprising nine log
smells and their associated facets, as well as five
direct causes of log smells and four direct consequences. To
  assess the applicability of the taxonomy beyond the surveyed
  literature, we also conducted a preliminary validation using a dataset of logging issues extracted from open-source projects.  We 
also defined a mapping between log smells (and their facets) and tools
that repair them, identifying 16 tools that address and
repair facets from six log smells.

The results of our study indicate that logging is
prone to issues and smells at all phases, affecting both the logging
code and the log files. As log smells represent logging issues that
hinder the quality of logging, potentially leading to more serious
consequences, it is crucial that practitioners and researchers 
understand, be aware of, and have methods for mitigating
them. Therefore, we believe that the presented taxonomy and overview of log smells, their causes, consequences, manifestations, and automated mitigation strategies provide a useful resource for researchers and practitioners seeking to understand, identify, and mitigate log smells.

As part of future work, we plan to further validate the taxonomy by conducting a larger-scale empirical Mining Software Repository (MSR) study. While the preliminary validation presented in this paper provided initial evidence that the taxonomy can be applied to real-world logging issues, the MSR study will investigate its occurrence and distribution across a broader set of software projects.
Furthermore, we intend to further explore specific log smells, 
provide a more detailed characterization of them, and empirically evaluate their
impact in the field. We also plan to develop 
tools for the log smells that currently lack automated solutions.

\section*{Acknowledgements}
This  research  was  funded  in  whole, or  in  part,  by  the
  Luxembourg  National  Research Fund (FNR), grant reference C22/IS/17373407/LOGODOR.

\end{document}